%% file: paper.tex
\documentclass[conference]{IEEEtran}

\input{preamble}

\input{figures/values}
\input{figures/values-infringing-clusters}
\input{figures/values-infringing-rate}
\input{figures/values-infringing-republished}
\input{figures/values-infringing-publishers}
\input{figures/values-survival}
\input{figures/values-kinds}
\input{figures/values-malware}

\begin{document}

\title{Did I Vet You Before? Assessing the Chrome Web Store Vetting Process through Browser Extension Similarity}

\author{
\IEEEauthorblockN{José Miguel Moreno}
\IEEEauthorblockA{%
\textit{Universidad Carlos III de Madrid}\\
Madrid, Spain\\
josemore@pa.uc3m.es}
\and
\IEEEauthorblockN{Narseo Vallina-Rodriguez}
\IEEEauthorblockA{%
\textit{IMDEA Networks Institute}\\
Madrid, Spain\\
narseo.vallina@imdea.org}
\and
\IEEEauthorblockN{Juan Tapiador}
\IEEEauthorblockA{%
\textit{Universidad Carlos III de Madrid}\\
Madrid, Spain\\
jestevez@inf.uc3m.es}
}

\maketitle

\begin{abstract}
Web browsers, particularly Google Chrome and other Chromium-based browsers, have
grown in popularity over the past decade, with browser extensions becoming an
integral part of their ecosystem.
These extensions can customize and enhance the user experience, providing
functionality that ranges from ad blockers to, more recently, AI assistants.
Given the ever-increasing importance of web browsers, distribution marketplaces
for extensions play a key role in keeping users safe by vetting submissions that
display abusive or malicious behavior.
In this paper, we characterize the prevalence of malware and other infringing
extensions in the Chrome Web Store (CWS), the largest distribution platform for
this type of software.
To do so, we introduce \textsc{SimExt}, a novel methodology for detecting
similarly behaving extensions that leverages static and dynamic analysis, Natural
Language Processing (NLP) and vector embeddings.
Our study reveals significant gaps in the CWS vetting process, as
\infringingRepublishedExtensionsP of infringing extensions are extremely similar
to previously vetted items, and these extensions take months or even years to
be removed.
By characterizing the top kinds of infringing extension, we find that
\kindsTotalTopNtesP are New Tab Extensions (NTEs) and raise some concerns about
the consistency of the vetting labels assigned by CWS analysts.
Our study also reveals that only \malwareMaliciousP of malware extensions flagged
by the CWS are detected as malicious by anti-malware engines, indicating a
concerning gap between the threat landscape seen by CWS moderators and the
detection capabilities of the threat intelligence community.
\end{abstract}

\input{sections/introduction}
\input{sections/background}
\input{sections/data-collection}
\input{sections/methodology}
\input{sections/rq1}
\input{sections/rq2}
\input{sections/rq3}
\input{sections/discussion}
\input{sections/related-work}
\input{sections/conclusion}

\bibliographystyle{IEEEtranS}
\bibliography{paper}

\appendices
\input{sections/appendix-serialization}
\input{sections/appendix-malware}
\input{sections/appendix-case-studies}

\end{document}

%% file: preamble.tex
\usepackage[hidelinks]{hyperref}
\usepackage{subcaption}
\usepackage{xcolor}
\usepackage{booktabs}
\usepackage{multirow}
\usepackage{makecell}
\usepackage{colortbl}
\usepackage{graphicx}

\usepackage{listings}
\definecolor{codeBackground}{RGB}{255,255,255}
\definecolor{codeComment}{RGB}{0,156,0}
\definecolor{codeKeyword}{RGB}{0,0,255}
\definecolor{codeLiteral}{RGB}{233,0,0}
\definecolor{codeString}{RGB}{233,0,0}
\definecolor{codeDefault}{RGB}{0,0,0}
\definecolor{codeNumbers}{RGB}{150,150,150}
\lstdefinelanguage{HTML5}{
  language=html,
  sensitive=true,
  alsodigit={.:;},
  alsoletter={<>=-},
  morecomment=[s]{<!-}{-->},
  tag=[s],
  escapeinside={<@}{@>},
  otherkeywords={
  >,
  <!DOCTYPE, <!doctype,
  </html, <html, <head, <title, </title, <style, </style, <link, </head, <meta, />,
  </body, <body,
  </div, <div, </div>,
  </p, <p, </p>, </h1, <h1,
  </script, <script,
  <canvas, /canvas>, <svg, <rect, <animateTransform, </rect>, </svg>, <video, <source, <iframe, </iframe>, </video>, <image, </image>, <header, </header, <article, </article
  },
  ndkeywords={
  =,
  charset=, src=, id=, width=, height=, style=, type=, rel=, href=,
  fill=, attributeName=, begin=, dur=, from=, to=, poster=, controls=, x=, y=, repeatCount=, xlink:href=,
  margin:, padding:, background-image:, border:, top:, left:, position:, width:, height:, margin-top:, margin-bottom:, font-size:, font-family:, line-height:,
  transform:, -moz-transform:, -webkit-transform:,
  animation:, -webkit-animation:,
  transition:,  transition-duration:, transition-property:, transition-timing-function:,
  }
}
\usepackage{xspace}
\newcommand\eg{e.g.,\xspace}
\newcommand\ie{i.e.,\xspace}
\newcommand\etal{et al.\xspace}

\usepackage{framed}
\newcommand{\takeaway}[1]{\begin{framed}\noindent{\textbf{Takeaway.} #1}\end{framed}}

%% file: figures/values.tex
\definecolor{blue}{HTML}{0c84ed}
\definecolor{orange}{HTML}{ff8811}
\definecolor{yellow}{HTML}{f8d90d}
\definecolor{green}{HTML}{38d04e}
\definecolor{red}{HTML}{ff0066}
\definecolor{purple}{HTML}{3b0086}

\newcommand\crawlingStart{March 2020\xspace}
\newcommand\crawlingEnd{March 2024\xspace}
\newcommand\dataStart{December 2009\xspace}
\newcommand\totalReleases{902,532\xspace}
\newcommand\totalExtensions{366,617\xspace}
\newcommand\totalThemes{53,132\xspace}
\newcommand\missingExtensions{28,405\xspace}
\newcommand\datasetSotaDelta{156k\xspace}

\newcommand\newExtsPerDay{106\xspace}
\newcommand\datasetDelta{74\%\xspace}
\newcommand\publishedMinusDelta{17\%\xspace}

\newcommand\deletedRatio{52\%\xspace}
\newcommand\vettedRatioOutOfDeleted{79\%\xspace}

\newcommand\tokensLenAvg{300\xspace}
\newcommand\tokensLenStd{228\xspace}
\newcommand\tokensBelowLimit{86\%\xspace}

%% file: figures/values-infringing-clusters.tex
\newcommand\infringingClusters{6,444\xspace}
\newcommand\infringingExtensions{100,957\xspace}
\newcommand\vettedExtensions{72,699\xspace}
\newcommand\vettedExtensionsP{72\%\xspace}
\newcommand\unpublishedExtensions{11,237\xspace}
\newcommand\unpublishedExtensionsP{11\%\xspace}
\newcommand\publishedExtensions{17,021\xspace}
\newcommand\publishedExtensionsP{17\%\xspace}
\newcommand\removedExtensionsInstalls{802M\xspace}
\newcommand\removedExtensionsInstallsMean{9.6k}
\newcommand\removedExtensionsInstallsStd{108k\xspace}
\newcommand\removedExtensionsInstallsTwentyFive{1\xspace}
\newcommand\removedExtensionsInstallsFifty{12\xspace}
\newcommand\removedExtensionsInstallsSeventyFive{146\xspace}
\newcommand\publishedExtensionsInstalls{172M\xspace}

\newcommand\publishedExtensionsInstallsTwentyFive{1\xspace}
\newcommand\publishedExtensionsInstallsFifty{17\xspace}
\newcommand\publishedExtensionsInstallsSeventyFive{235\xspace}

%% file: figures/values-infringing-rate.tex
\newcommand\infringingRateFull{1,397\xspace}
\newcommand\infringingRateFullP{22\%\xspace}
\newcommand\infringingRateMean{0.59\xspace}
\newcommand\infringingRateTwentyFive{0.33\xspace}
\newcommand\infringingRateSeventyFive{0.88\xspace}

%% file: figures/values-infringing-republished.tex
\newcommand\infringingRepublishedClustersP{92\%\xspace}
\newcommand\infringingRepublishedExtensions{86k\xspace}
\newcommand\infringingRepublishedExtensionsP{86\%\xspace}

%% file: figures/values-infringing-publishers.tex
\newcommand\infringingRepeatOffenders{6,156\xspace}
\newcommand\infringingRepeatOffendersWithPublished{663\xspace}
\newcommand\infringingRepeatOffendersWithPublishedP{11\%\xspace}
\newcommand\infringingPublishedByRepeatOffenders{1,896\xspace}
\newcommand\infringingPublishedByRepeatOffendersP{11\%\xspace}

%% file: figures/values-survival.tex
\newcommand\survivalAllSize{100k\xspace}
\newcommand\survivalSampleSize{1,000\xspace}

\newcommand\survivalMoreThanOneYearP{59\%\xspace}
\newcommand\survivalMoreThanTenYears{23\xspace}

\newcommand\survivalLessThanOneMonthP{13\%\xspace}

\newcommand\survivalLessThanOneWeekP{4\%\xspace}
\newcommand\survivalKmMedian{581\xspace}
\newcommand\survivalKmProbMoreTenYears{6\%\xspace}
\newcommand\survivalKmPolicyViolationMedian{743\xspace}
\newcommand\survivalKmMinorPolicyViolationMedian{443\xspace}
\newcommand\survivalKmMalwareMedian{282\xspace}

%% file: figures/values-kinds.tex
\newcommand\kindsTopClusters{65\xspace}
\newcommand\kindsTopNteClusters{15\xspace}
\newcommand\kindsTotalExtensions{44k\xspace}
\newcommand\kindsTotalPublishers{6.2k\xspace}
\newcommand\kindsTotalTopNtes{37k\xspace}
\newcommand\kindsTotalTopNtesP{83\%\xspace}
\newcommand\kindsTotalTopPublishedNtes{55\xspace}
\newcommand\kindsTotalPublishedNtes{571\xspace}

%% file: figures/values-malware.tex
\newcommand\malwareExtensions{5,647\xspace}
\newcommand\malwareNotFound{5,377\xspace}
\newcommand\malwareNotFoundP{95\%\xspace}

\newcommand\malwareCleanP{3\%\xspace}
\newcommand\malwareMalicious{83\xspace}
\newcommand\malwareMaliciousP{1\%\xspace}
\newcommand\malwareMaliciousClusters{7\xspace}
\newcommand\malwareMaliciousOutliers{68\xspace}

%% file: sections/introduction.tex
\section{Introduction}
\label{sec:introduction}

Over the past decade, web browsers have become an indispensable tool for both
desktop and mobile users.
As traditional applications migrate to web-based services, users spend more time in
their browsers, increasing the need for extended and more customization features.
Chromium-based browsers---notably, Google Chrome---allow this customization
through browser extensions:
third-party programs built with web technologies that interact with the browser
to add or modify features, and enhance the user experience.
Some popular types of extensions include ad blockers, grammar checkers, password
managers, and, more recently, AI assistants.

The undisputed marketplace for browser extensions for Chro\-mi\-um browsers
is the Chrome Web Store (CWS), Google's official marketplace which launched
back in 2011~\cite{google-initial-cws-announcement}.
The CWS plays a vital role in the browser ecosystem as many users inherently
trust the extensions included on its catalog.
Similar to the case of the Google Play Store for Android apps, the CWS implements
vetting processes to detect and mitigate the risk of harmful and deceptive
extensions being distributed through the official store.
Extension publishers must adhere to the CWS Program Policies, which govern what
content can be uploaded and distributed~\cite{chrome-policies}.
These policies forbid not only the publication of harmful content, but also mandate
that extensions must disclose their behavior, implement a minimum of functionality,
and not duplicate existing content, among other requirements.
If an infringing extension (\ie one that does not abide by the CWS policies) is
published, Google may delist it at its discretion during the vetting process
and, in some cases, even remove it from browsers that have it installed.

In the past, the CWS has been under scrutiny from security experts and researchers.
Various reports and research studies identified infringing content being distributed
on the CWS, focusing mostly on potentially harmful
extensions~\cite{report-palant-malicious-extensions, report-kaspersky-dangerous-extensions,
report-cashback-extension-killer} and in methods to detect their
presence~\cite{hulk, trends-and-lessons, i-spy-with-my-little-eye, youve-changed,
no-signal-left-to-chance}.
While there is evidence of Google proactively vetting extensions from its official
store~\cite{trends-and-lessons}, the effectiveness of the vetting process at
detecting infringing content remains unexplored.
Recent research has focused on characterizing the content of the store and its
trends~\cite{what-is-in-the-cws}, including the presence of extensions with
vulnerabilities or a similar code base.
However, a knowledge gap remains when it comes to understanding Google's policing
of the platform.
Characterizing the challenges that the CWS vetting process faces in practice
can help identify its pitfalls and improve its effectiveness.
To fill this gap, this paper aims to answer the following research questions:

\begin{itemize}[
  \setlabelwidth{RQX.}
  \settowidth{\labelindent}{}
  \settowidth{\labelsep}{X}
]
  \item[\textbf{RQ1.}] How effective is the CWS vetting process at detecting and rapidly removing infringing extensions?
  \item[\textbf{RQ2.}] What are the main kinds and the key features of infringing extensions uploaded to the CWS?
  \item[\textbf{RQ3.}] What malware families target the CWS and what threat
  intelligence is available about them?
\end{itemize}

To answer these questions, we develop the scalable data collection and hybrid
analysis pipeline for browser extensions shown in Figure~\ref{fig:methodology-pipeline}.
We compile a dataset of \totalExtensions extensions downloaded from the CWS
during 4 years.
We enrich this dataset with vetting labels assigned by Google moderators to
removed CWS items.
These labels indicate which extensions have been taken down from the store and
for what reason, thus providing ground truth about the vetting process.
To the best of our knowledge, this is the first work
that studies the CWS vetting process using ground truth about removed content.
We then design and develop \textsc{SimExt}, a novel methodology for finding
similarly behaving extensions that leverages $(i)$ static and dynamic analysis
for extracting syntactic and behavioral features; and $(ii)$ Natural Language
Processing (NLP) and vector embeddings for computing similarity.
To address the existing technical gap in feature extraction from browser
extensions, we develop and release Fakeium~\cite{fakeium}, an open source
dynamic analysis sandbox that follows an implementation inspired by concolic
execution to elicit behaviors and overcome the limitations of static
analysis and fuzzy-hash-based methods~\cite{what-is-in-the-cws}.
Fakeium is very scalable and can process an extension in mere
seconds, which makes it suitable for large-scale measurements such as the
one conducted in this work.
To generate the vector embeddings, we use a novel approach based on Zero-Short
Learning (ZSL) which, unlike previous work, does not require retraining the model
when new malicious extensions are uploaded to the store~\cite{trends-and-lessons,
i-spy-with-my-little-eye, youve-changed, no-signal-left-to-chance}.

Using \textsc{SimExt}, we perform a large-scale behavioral cluster analysis to
find published extensions that are very similar to vetted ones, as these are
natural candidates for being infringing items that were overlooked by the vetting
process.
Equipped with this pipeline, we make the following novel contributions about the
CWS vetting process and its effectiveness:

\begin{enumerate}[
  \setlabelwidth{X}
  \settowidth{\labelindent}{}
  \settowidth{\labelsep}{X}
]
\item We find \publishedExtensions potentially infringing extensions that are still
published on the CWS.
Our pipeline flags these extensions as extremely similar to known infringing
extensions previously taken down from the store.
Interestingly, \infringingRepublishedExtensionsP of these infringing items are
republished extensions, \ie extensions that are either identical or extremely
similar to items that were previously taken down, suggesting that the CWS vetting
process does not adequately learn from experience.
We also find that \infringingPublishedByRepeatOffendersP of the infringing
extensions that remain published at the end of our crawl come from repeat offenders
(publishers with known vetted items), suggesting that the CWS does not suspend
these accounts according to their own policies.
We also conduct a survival analysis for infringing extensions.
Our results reveal that the process of identifying infringing extensions is
excessively slow, taking months or even years from the publication of an
infringing version to its removal from the store.

\item To validate our findings, we analyze extensions from the top infringing
clusters with the most extensions and manually assigned them a type or kind of
content.
We find that \kindsTotalTopNtesP of the extensions in these clusters are New Tab
Extensions (NTEs), which are items that override the webpage that loads when a
new tab is opened.
The cluster analysis also reveals the presence of spam and no content extensions,
both vetted and still published in the CWS.
In order to better understand the labeling process, we explore the relationship
between the CWS vetting labels and the behavior of an extension.
Our analysis finds that labels are somewhat inconsistent and unrelated to
features or behaviors, especially in the case of NTEs.

\item We scan the \malwareExtensions extensions labeled as malware by the CWS in
VirusTotal and compare the results.
Remarkably, \malwareNotFoundP of malware-labeled samples were not indexed by
VirusTotal and \malwareCleanP were known but have zero detections.
This finding suggests that the majority of malicious extensions detected in the
CWS are unknown to the threat intelligence community, and that for others there
is a significant discrepancy between the criteria of CWS moderators and the
detection capabilities of malware detection engines.
Analysis of the malicious extensions for which we obtain detection reports reveals
that the malware lineage analysis conducted by security vendors is very poor,
with the family label missing in most cases and being very general in others.

\end{enumerate}

\vspace{1mm}
\noindent\textbf{Disclosure.}\xspace
We reported to Google a subset of 180 infringing extensions found by our
pipeline.
All of them were manually reviewed to confirm the presence of a policy violation.
Most importantly, this set includes 23 no content extensions published from
corporate Google accounts, all of which were removed from the store shortly after
our report.
As of this writing, we have not received an official response.

\vspace{1mm}
\noindent\textbf{Tool and Research Artifacts.}\xspace
We release the source code of Fakeium, an instrumented JavaScript sandbox that
can be used to extract API calls from browser extensions~\cite{fakeium}.
We also provide the list of extension pairs used as ground truth for evaluation, and
metadata and high-dimensional embeddings for infringing extensions found by our
pipeline at \url{https://zenodo.org/records/10977708}.

%% file: sections/background.tex
\section{Background}
\label{sec:background}

\begin{figure*}[t]
  \centering
  \includegraphics[width=\textwidth]{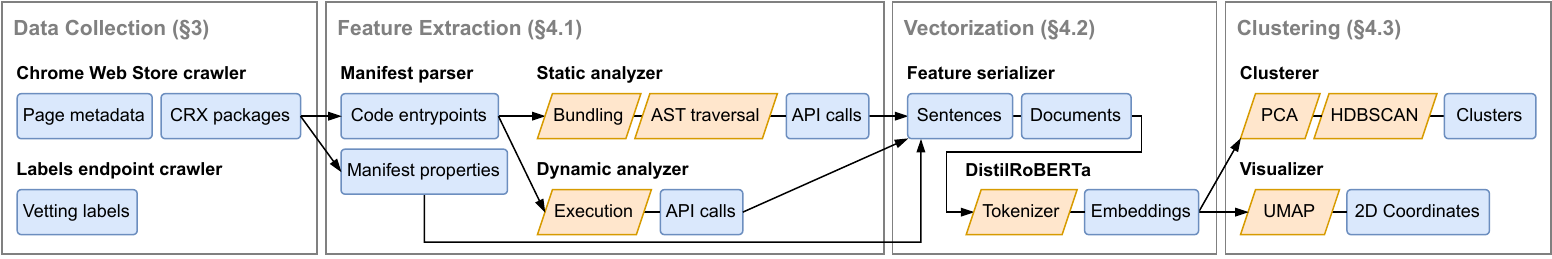}
  \caption{Data collection and analysis methodology pipeline.}
  \label{fig:methodology-pipeline}
\end{figure*}

This section provides background on browser extensions and the CWS
publishing ecosystem.

\subsection{Browser Extensions}
Browser extensions are third-party programs that enhance or modify the
functionality of the browser, with ad blockers being one of their most
recognizable and widely-used examples.
Extensions are written using a combination of HTML, CSS and JavaScript, and
may contain other resources commonly found in web applications, such as
images or fonts~\cite{mdn-browser-extensions}.
Extensions are packaged as CRX files, bundling all the source code and assets
needed to run them into a single file, thus easing distribution.
CRX packages are regular ZIP archives with an additional header that
ensures the integrity and authenticity of its contents~\cite{crx3-design-doc}.

As in the case of other popular platforms like Android,
all extensions have a mandatory \texttt{manifest.json} file that dictates
their functionality and behavior.
This manifest is a JSON document that specifies basic metadata about the
extension, like its name and version.
Optionally, it may also declare permissions, service workers (which run
code in the background), content scripts (which run code in open tabs), and
overridden pages (\eg to modify the default URL when opening a new tab),
among other elements~\cite{chrome-manifest}.
Two popular types of extensions are themes and New Tab Extensions (NTEs).
A theme is a type of extension that changes the way the browser looks and
does not contain any source code, \ie all theme configuration is in the
manifest itself.
An NTE is an extension that changes the default webpage that will load every
time a new tab is opened.

Extensions communicate with the browser using Web APIs~\cite{mdn-web-apis} and
Extension APIs~\cite{chrome-api-reference}.
These APIs provide extensions with a broad range of powerful capabilities,
allowing them to access and manipulate the Document Object Model (DOM)
of webpages, and to access geolocation data, browsing history, and other
sensitive information stored in the browser.
Some of these APIs are restricted by default for security reasons, and can
only be accessed by declaring the appropriate permissions in the manifest.

\subsection{CWS Program Policies}
CRX packages can be published on the Chrome Web Store (CWS), Google's
official distribution platform for browser extensions that run on Google
Chrome~\cite{chrome-web-store}.
Users of other Chromium-based brows\-ers, such as Microsoft Edge, Brave, or Opera,
can also install extensions from this source as they run the same compatible
engine.
In fact, Brave even recommends that users download extensions from this platform
in the absence of their own official store~\cite{brave-installing-extensions}.
As such, policing the content that is uploaded to the CWS is of great
importance because of the impact that harmful extensions can have
given the large audience of the store.

The CWS requires all developers to comply with a set of policies that include
security and privacy guidelines.
These policies prohibit the publication of extensions with a malicious purpose
and extensions that facilitate unauthorized access to restricted or copyrighted
content~\cite{chrome-policies-prohibited}.
They also ask developers not to obfuscate code or conceal functionality, to
request the minimum set of permissions needed, and forbid submitting multiple
extensions that provide duplicate or very little
functionality~\cite{chrome-policies-readability, chrome-policies-spam, chrome-policies-minimum-functionality}.
According to Google's policies, repeated violations of these policies will result
in the suspension of the publisher account~\cite{chrome-policies-repeat-abuse}.

%% file: sections/data-collection.tex
\section{Data Collection}
\label{sec:data-collection}

We implemented a purpose-specific crawler to download the resources shown in
Figure~\ref{fig:methodology-pipeline}, namely CRX packages and their associated
metadata, from Google's Chrome Web Store~\cite{chrome-web-store}.
Our crawling strategy consists of three steps.
First, we prepare an initial set of extensions from the store's sitemap, the
reels of popular items from each category, and the featured extensions from the
home page.
We then visit each extension page, extract metadata (\eg name, description,
number of installs, last modification date) and download its associated assets
(\ie CRX package, icon and tile).
Finally, we look for non-visited extensions appearing as ``related'' or
``recommended'' in the store page of each visited extension and add them to the
queue of items to crawl.
Previous efforts demonstrate the effectiveness of this
approach~\cite{doublex, fingerprinting-in-style}, which takes inspiration from
strategies successfully used to crawl the
Google Play Store~\cite{beyond-google-play, rise-of-the-planet-of-the-apps}.

We run the crawler daily to detect new and removed extensions, and keep track of
changes being pushed to the store such as publishers uploading an updated version
of an existing extension.
For simplicity, we set the visitor's country code to the United States when
crawling the store.
Although developers can opt-out of publishing extensions in certain countries or
even entire regions~\cite{chrome-prepare-to-publish}, this decision should not
limit our coverage as we can still download any non-private CRX package regardless,
and because the used sitemap contains publicly listed extensions worldwide.
We only run our data collection pipeline once a day and rely on the HTTP caching
headers provided by CWS to responsibly limit our impact on Google's servers.
We note that in October 2023, we transitioned to crawling the new version of the
CWS, coinciding with its first public release~\cite{google-launched-redesigned-cws}.
We made this decision in preparation for the announced retirement of the previous
version in January 2024.
While we experienced no issues during this migration, we also kept crawling the
legacy store for a few more weeks in the event something went wrong.

In addition to crawling CRX packages and store metadata, we also fetch vetting
labels issued by Google from a separate endpoint.
These labels indicate whether an extension was taken down from the CWS instead of
willingly being unpublished by its developer, as well as the reason behind the
takedown (\eg flagged as malware), if any.
Vetting labels are used by Google Chrome since version 117 to protect users of
vetted extensions by proactively disabling malware and displaying warnings in
the browser extensions page~\cite{extension-safety-hub}.
As far as we can tell, this is the first study to crawl and analyze this data.

\subsection{Dataset}
We crawled the CWS for 4 years from \crawlingStart to \crawlingEnd.
We compiled a historical dataset of \totalExtensions extensions, containing
\totalReleases different updates (or extension \textit{versions}) that were
available for download at some point in the store.
Some of the extensions in our datasets were published as early as
\dataStart.\footnote{
  While the CWS was officially launched in 2011, it existed prior to
  that date as evidenced by \url{https://narkive.com/bEuDVJqN}
}
Of all these versions, we lack the CRX package for \missingExtensions paid
extensions that did not change their pricing model after payments were deprecated
in the CWS in mid-2020~\cite{chrome-payments-deprecation}.
Our dataset also contains \totalThemes browser themes, which are distributed
in the same form as extensions but do not have the ability to execute code.
To the best of our knowledge, this is the most comprehensive dataset of browser
extensions to-date, with over \datasetSotaDelta more extensions than the previous
large-scale work that crawled the CWS~\cite{youve-changed}.

\begin{figure}[t]
  \centering
  \includegraphics[width=0.9\columnwidth]{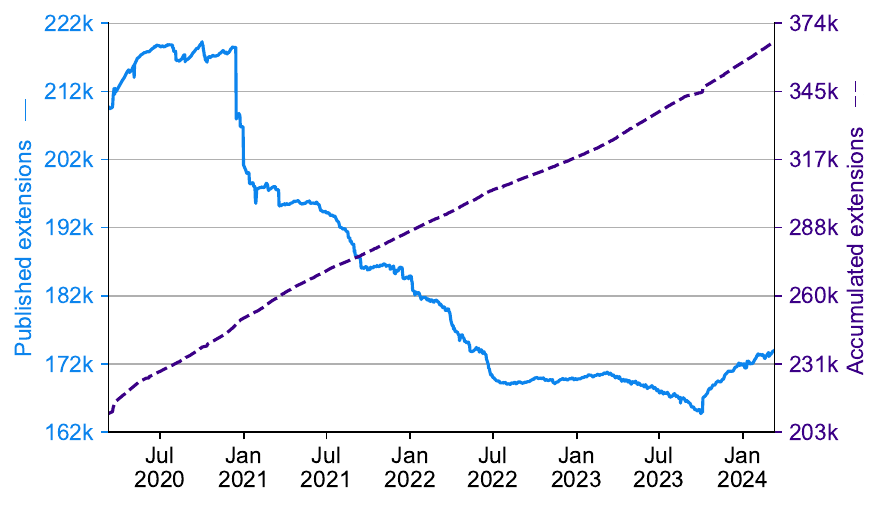}
  \caption{
    Daily volume of published extensions (solid, left axis)
    and accumulated dataset size (dashed, right axis).
  }
  \label{fig:dataset-volume}
\end{figure}

Figure~\ref{fig:dataset-volume} shows the change over time of the number of
currently published extensions in the CWS (solid line) versus the total size
of our dataset (dashed line).
We observe that our dataset grows by an average of \newExtsPerDay new extensions
per day, with this rate remaining fairly consistent throughout the 4 years of our
study.
Yet, the number of published extensions in the store has decreased by
\publishedMinusDelta since \crawlingStart, experiencing a noticeable drop in
December 2020 and following a downward trend since then.
This contrasts with the \datasetDelta increase of our dataset in the same time
period.
Without further analysis of the dataset, this growth suggests a very dynamic
ecosystem with a high rotation of extensions entering and leaving the store.
This rotation could either be attributed to developers deliberately publishing
and shortly thereafter unpublishing their extensions, or Google taking down a
large number of malware and other abusive content as part of its vetting process.

\vspace{1mm}
\noindent\textbf{Vetting Labels.}\xspace
Using the vetting labels that we crawl for removed extensions, we can infer whether
they were unpublished by the developer or taken down by Google.
Vetted extensions have a type of either \textit{Malware},
\textit{Policy Violation}, or \textit{Minor Policy Violation}, whereas items that
were unpublished at the request of their developers have a type of \textit{None}.

\begin{figure}[t]
  \centering
  \includegraphics[width=0.9\columnwidth]{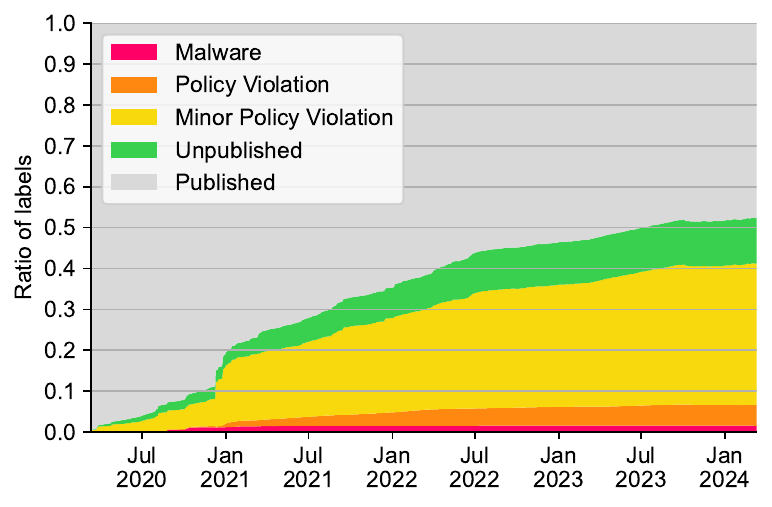}
  \caption{Daily ratio of vetting labels provided by CWS.}
  \label{fig:dataset-labels}
\end{figure}

Figure~\ref{fig:dataset-labels} shows the ratio of extension labels found in our
dataset at a given point in time since we first started crawling.
As of \crawlingEnd, \deletedRatio of all the extensions we collected have been
removed from the store.
More interestingly, \vettedRatioOutOfDeleted of the removed items are infringing
extensions that were taken down.
This preliminary characterization of the dataset motivates us to delve deeper
into the dynamics of extension removals and the vetting process of the CWS.

%% file: sections/methodology.tex
\section{SimExt: A Methodology for Finding Similar Extensions}
\label{sec:methodology}

While developing a fully automated vetting process is very challenging, having
an automatic approach to detect similar extensions at scale is valuable to
assist analysts, as it can be used to find items that resemble previously vetted
ones or clones to currently published content.
Our key working hypothesis is that if an extension is taken down from the
store because it violates a program policy, then other extensions that are
highly similar are candidates to be vetted as well.
Even in those cases where the behavior of an extension is not necessary
harmful, finding clones is still valuable given the CWS ban on repetitive
content~\cite{chrome-policies-spam}.
We acknowledge that legitimate extensions cloned by malicious actors are an
exception to this rule.
For example, the vetting of malicious extensions reusing MetaMask's codebase
should not result in MetaMask being taken down.
We consider that these cases should be manually filtered out, and that a
tool that provides detailed similarity reports between two extensions can
be extremely helpful to assist the store vetting process.
Section~\ref{sec:rq2} provides examples of published infringing extensions
found by our tool to support this claim.

Since there is currently no tool that identifies similar extensions for a given
input item, we designed and built \textsc{SimExt}, our own state-of-the-art
solution for querying browser extensions based on similarity.
We design it as a multi-step pipeline to automatically extract meaningful
behavioral features from all extensions in our dataset and then group them into
clusters based on their similarity.
This approach is different from previous work that classifies extensions as either
benign or malicious~\cite{trends-and-lessons, i-spy-with-my-little-eye, youve-changed}.
Moreover, to the best of our knowledge, our work is the first one to leverage
Natural Language Processing (NLP) and vector embeddings to effectively and
accurately find similar extensions at scale.
Figure~\ref{fig:methodology-pipeline} provides a summary of the pipeline steps
described below.

\subsection{Feature Extraction}
\textsc{SimExt} uses both static and dynamic analysis to extract manifest
properties and API calls to determine the structure and behavior of an
extension.

\vspace{1mm}
\noindent\textbf{Manifest.}\xspace
We flatten the properties found in the extension manifest to convert them from a
nested JSON object to key-value pairs.
We exclude the \texttt{author}, \texttt{name}, \texttt{short\_name},
\texttt{description}, \texttt{version}, \texttt{key} and \texttt{update\_url}
properties as they typically do not provide reliable information about the behavior
of an extension and, in the worst case, could help the attacker to trick the model
into giving a different embedding to the extension.
We also identify and enumerate all code entrypoints found in the manifest.
Per Google's documentation~\cite{chrome-manifest}, we consider any of the
following items to be valid entrypoints:
$(i)$ content scripts;
$(ii)$ background pages or service workers;
$(iii)$ popup pages, \ie browser actions;
and $(iv)$ overridden, DevTools, side panel, or options pages.

\vspace{1mm}
\noindent\textbf{Static Analysis.}\xspace
We statically analyze the JavaScript code of entrypoints.
We use esbuild~\cite{esbuild} for module resolution, bundling, tree shaking
(\ie dead code removal) and minification.
We then use Babel~\cite{babeljs}, a well-maintained and widely used tool in the
modern web ecosystem, to generate the Abstract Syntax Tree (AST) of the resulting
bundle.
Although previous work has relied on Esprima~\cite{esprima} for this
purpose~\cite{doublex, youve-changed, dangers-of-human-touch, empoweb, hardening-browser-extensions},
we decided to use a more modern alternative given that Esprima has not received any
major updates since 2018.
Lastly, we traverse the AST to extract the list of API calls as relevant features
to model the behavior of the extension.
We define an \textit{API call} as an invocation to an Extension
API~\cite{chrome-api-reference} from the \texttt{chrome} or \texttt{browser}
object, or to a Web API~\cite{mdn-web-apis} within the \texttt{navigator} object.
In particular, we use the Babel Traverse module and the
\texttt{ReferencedIdentifier} and \texttt{BinaryExpression} nodes to find the
names of global variables (\eg with no bindings except for the global scope)
being invoked in a program.

\vspace{1mm}
\noindent\textbf{Dynamic Analysis.}\xspace
To overcome the limitations of static analysis and detect API calls we might
otherwise miss (\eg code being executed in a call to \texttt{eval}), we complement
our feature extraction pipeline with a dynamic analysis stage.
We develop Fakeium, our own sandbox environment based on
isolated-vm~\cite{isolated-vm} to safely run untrusted JavaScript code.
Fakeium intentionally lacks the webpage DOM and Web and Browser APIs.
Instead, it works by mocking any objects accessed by the extension at runtime
to prevent it from crashing for as long as possible.
This mocking is performed by hijacking the global scope using a \texttt{Proxy}
object~\cite{mdn-proxy}.
We monitor calls to mocked objects and log API calls as in static analysis.
To ensure that we cover as many code paths as possible, we follow an approach
inspired by concolic execution, where we invoke all callbacks and functions found
during the analysis, whilst still respecting the original conditional predicates.
This approach enables Fakeium to run without the need for an automation tool or
\textit{monkey} to trigger behaviors.
While not as accurate as running the extension in an actual web browser, this
approach is more scalable, taking just a few seconds instead of minutes to analyze
an extension.
In addition, this environment relies on the same JavaScript engine that Chromium
uses~\cite{v8}, guaranteeing accurate output for code snippets that depend only
on the ECMAScript specification (\ie without invoking APIs that will be mocked
by our sandbox).
Fakeium is released under the MIT license and distributed as a Node.js
package~\cite{fakeium}.

\subsection{Vectorization}
\label{sec:methodology:vectorization}
We apply Natural Language Processing (NLP) to generate a vector embedding
for each extension based on its features.
First, we serialize manifest properties and API calls into text \textit{sentences}
that are sorted alphabetically and concatenated with a semicolon to produce
\textit{documents}.
Specifically, we remove all punctuation and normalize word casing for manifest
keys and API calls, and perform a laxer transformation for manifest values.
We merge similar manifest keys together and limit the number of values per group
to avoid extremely long documents.
Appendix~\ref{sec:appendix-serialization} provides an example of these feature
serialization techniques for reference.

Inspired by Zero-Shot Learning (ZSL)~\cite{train-once-test-anywhere}, we use
Sentence-Transformers~\cite{sentence-transformers} with the ``all-distilroberta-v1''
variant of the DistilRoBERTa model\footnote{
  See \url{https://huggingface.co/sentence-transformers/all-distilroberta-v1}
} to compute 768-dimensional embeddings of the generated documents.
DistilRoBERTa is a general purpose model based on the BERT framework that was
trained on a large and diverse corpus of English texts~\cite{roberta}.
It uses knowledge distillation to improve performance while retaining a high
accuracy~\cite{distilbert}.
We found this model to be a good fit for our needs as it achieves a good balance
between sentence performance and encoding speed.\footnote{
  See comparison at \url{https://www.sbert.net/docs/pretrained_models.html}
}
Using the default tokenizer provided with the model, the average length of our
documents is \tokensLenAvg tokens, with a standard deviation of \tokensLenStd
tokens.
While some inputs are truncated after exceeding the maximum sequence length of
512 tokens set by the model, this number is fairly small as \tokensBelowLimit of
the documents fit within the limit.

\subsection{Clustering}
We employ density-based clustering to group together extensions with a very
similar behavior based on their embeddings, and to filter out outliers that do
not belong to any cluster.
Prior to the clustering, we use Principal Component Analysis (PCA) to reduce the
dimensionality of the standardized embeddings for performance while retaining
95\% of the amount of variance.
This results in PCA deciding to keep 161 components.
Since DistilRoBERTa already outputs normalized embeddings, we skip the
standardization step.
Then, we use Hierarchical Density-Based Spatial Clustering of Applications with
Noise (HDBSCAN)~\cite{hdbscan}.
We choose HDBSCAN over other clustering algorithms like K-Means as it does not
require knowing the number of clusters beforehand.
We pick a minimum cluster size of 5 data points and lower the minimum samples
parameter to 2.
This configuration should discard points with low stability (\ie outliers) while
still allowing us to create clusters in less dense areas.

\vspace{1mm}
\noindent\textbf{Visualization.}\xspace
To facilitate the visualization of the extensions in our dataset and their
similarities, we use Uniform Manifold Approximation and Projection
(UMAP)~\cite{umap} to project the high-dimensional space formed by the
standardized extension embeddings onto a 2D plane.
This step does not affect the results of the clustering process, as we rely
solely on HDBSCAN for assigning cluster IDs to extensions.

\subsection{Evaluation}
To validate the performance of \textsc{SimExt}, we use a dataset of similar and
different extensions and check whether the pipeline correctly puts similar pairs
in the same cluster and different pairs in different clusters.

\vspace{1mm}
\noindent\textbf{Ground Truth.}\xspace
Given the lack of previous work in this area, we could not find a dataset
of similar CRX files, nor a tool to facilitate this process.
Therefore, we create our own ground truth by listing popular extensions in the
CWS and using the embeddings generated by our pipeline to find similar versions
of these items.
Additionally, we take advantage of developers publishing the same extension across
stores to add them to the list of similar pairs.
That is, we manually look for samples of featured CWS extensions that are also
published on Edge Add-ons~\cite{edge-addons} or Opera add-ons~\cite{opera-addons},
as we expect those to be identical.
We build pairs of different extensions by looking for items that offer the same
functionality but implement it differently.
For example, uBlock Origin and Privacy Badger make a pair of different ad blockers.

We consider two items to be similar if they meet all of the following criteria:
$(i)$ their manifests have an overlap of $\geq$90\% of their keys;
$(ii)$ their manifests have unique values in common, such as file paths or
localized strings;
$(iii)$ their file trees have $\geq$50\% of paths in common, excluding localized
\texttt{messages.json} files; and
$(iv)$ they share source code files that are identical after beautification,
excluding third-party libraries.

For a fairer evaluation, we complement the ground truth with manually-reviewed
pairs of similar and different extensions taken from random clusters obtained
after building the pipeline.
We provide the resulting list of pairs as an artifact for reproducibility
(see Section~\ref{sec:introduction}).

\vspace{1mm}
\noindent\textbf{Results.}\xspace
Using the process described above, we managed to put together a ground truth
dataset of 110 pairs of similar extension versions, and another 110
pairs of different ones (220 unique pairs in total).
When validating against the comprised ground truth, \textsc{SimExt} shows an
accuracy of 83.2\%, with a precision of 85.4\% and a recall of 80.0\%.
We consider these results adequate for the task:
as the first work to use embeddings for clustering browser extensions, the high
accuracy obtained gives us confidence in the proposed novel methodology and
positions us well to conduct the first examination of the CWS vetting process.

%% file: sections/rq1.tex
\section{Effectiveness of the Vetting Process}
\label{sec:rq1}

This section examines the effectiveness of the CWS vetting process at identifying
content that fails to comply with store policies and the time that offending items
remain published before they are eventually taken down (RQ1).
To answer this question, we take the most recent version for each extension
(including removed extensions) as of the end of our crawl in \crawlingEnd.

\subsection{Infringing Extensions Detection}
As discussed in Section~\ref{sec:methodology}, we assume that extensions that are
very similar to vetted extensions are most likely in violation of store policies
and should likely be taken down.
We acknowledge that this is a strong assumption, and we validate it in
Section~\ref{sec:rq2} by performing a manual analysis of the clusters and providing
compelling examples of nearly identical extensions found by \textsc{SimExt}.
To benchmark how good the CWS is at finding these potentially infringing extensions,
we quantify how many extensions that are still published as of the latest
snapshot of our dataset belong to a vetted cluster.
We define an \textit{infringing cluster} as a set of 2 or more similarly behaving
extensions in which at least one of them has been taken down by Google (\ie has
been vetted) according to the the labels they provide, as presented in
Section~\ref{sec:data-collection}.

\begin{figure}[t!]
  \centering
  \begin{subfigure}[b]{\columnwidth}
    \centering
    \includegraphics[width=0.9\linewidth]{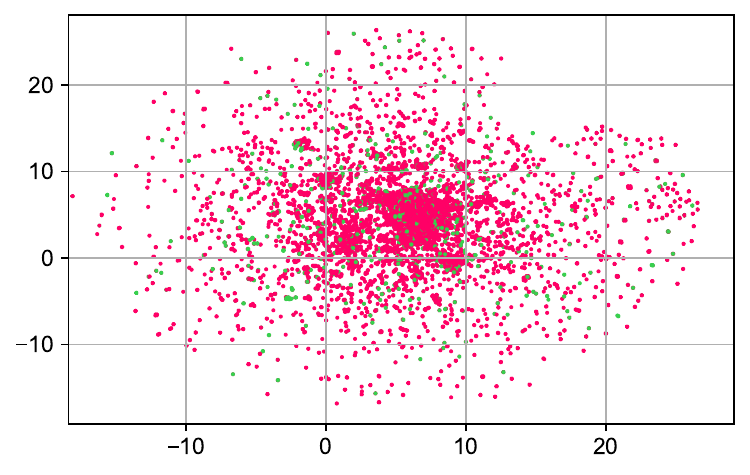}
    \caption{Removed extensions}
    \label{fig:infringing-scatter:removed}
  \end{subfigure}
  \begin{subfigure}[b]{\columnwidth}
    \centering
    \includegraphics[width=0.9\linewidth]{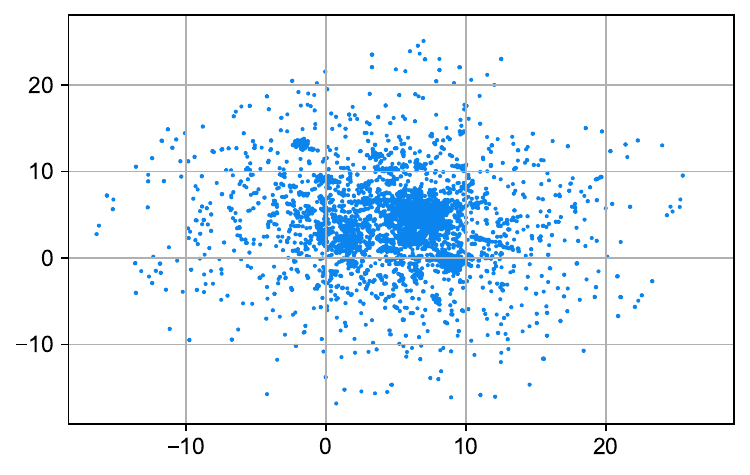}
    \caption{Published extensions}
    \label{fig:infringing-scatter:published}
  \end{subfigure}
  \caption{
    Scatter plots of extensions belonging to an infringing cluster.
    Taken down (vetted) extensions in \textcolor{red}{red},
    unpublished by the developer in \textcolor{green}{green},
    and still published in \textcolor{blue}{blue}.
  }
  \label{fig:infringing-scatter}
\end{figure}
Figure~\ref{fig:infringing-scatter} shows two-dimensional projections of the
embeddings of extensions found in infringing clusters.
In total, our tool finds \infringingClusters infringing clusters comprised of
\infringingExtensions extensions.
These clusters contain \vettedExtensions extensions that were taken down
(\vettedExtensionsP), and an additional smaller---yet significant---set of
\unpublishedExtensions unpublished extensions (\unpublishedExtensionsP).
According to CWS labels, the aforementioned were voluntarily unpublished by their
developers without Google raising a policy violation despite being similar to
vetted extensions.
In Section~\ref{sec:discussion} we discuss why we believe publishers might
prefer to intentionally remove their extensions over receiving a takedown by Google.
Lastly, our pipeline finds \publishedExtensions published extensions that account
for the remaining \publishedExtensionsP of infringing clusters.
Overall, these numbers are reasonably good and show that the CWS has some
capability to detect infringing extensions.
Nevertheless, we reiterate that the number of published infringing items is an
estimate derived from our pipeline and the actual amount may vary.

\vspace{1mm}
\noindent\textbf{Impact.}\xspace
We measure the impact of infringing extensions based on the number of active user
installs, using the most recent count available.
Removed extensions have a cumulative sum of \removedExtensionsInstalls users,
whereas published extensions have \publishedExtensionsInstalls users in total.
The mean for removed items stands at
\removedExtensionsInstallsMean$\pm$\removedExtensionsInstallsStd users,
indicating substantial variability, with 25\%, 50\%, and 75\% of installs lying
below \removedExtensionsInstallsTwentyFive,
\removedExtensionsInstallsFifty,
and \removedExtensionsInstallsSeventyFive, respectively.
For published extensions, this distribution is slightly higher, with users
falling beneath \publishedExtensionsInstallsTwentyFive,
\publishedExtensionsInstallsFifty, and \publishedExtensionsInstallsSeventyFive
for the same ranges above.
As such, we find that the number of users does not meaningfully change between
removed and published but potentially infringing extensions.
This suggests that Google removes items regardless of their popularity.

\vspace{1mm}
\noindent\textbf{Detection Rate.}\xspace
To assess how effective the CWS is at detecting infringing extensions across
clusters of different size, we define the the \textit{detection rate} for a
cluster as the number of vetted extensions in the cluster
divided by its size.
Ideally, a perfect similarity tool with a flawless vetting process should provide
100\% rate for all infringing clusters, implying that all policy-violating items
are taken down by the store operator.
Instead, our tool finds that only \infringingRateFull clusters
(\infringingRateFullP) have a perfect rate.
On average, infringing clusters have a detection rate of \infringingRateMean,
widely ranging from \infringingRateTwentyFive to \infringingRateSeventyFive from
the lower to the upper quartile.

\vspace{1mm}
\noindent\textbf{Republished Extensions.}\xspace
In terms of effectiveness at finding infringing extensions, we also examine
whether the CWS learns from experience and blocks the publication of extensions
that are similar to previously vetted ones.
To answer this, we take the earliest vetted extension from each infringing cluster
and count the number of \textit{republished extensions}, \ie similar items that
were published at a later date.
According to our pipeline, \infringingRepublishedClustersP of infringing clusters
contain extensions that were published after the fact.
Furthermore, republished extensions make up to \infringingRepublishedExtensionsP
of all infringing extensions, or \infringingRepublishedExtensions items in total.
These results ultimately suggest that CWS lacks rules for blocking extensions
similar to known vetted items.

\vspace{1mm}
\noindent\textbf{Repeat Offenders.}\xspace
We find evidence of \infringingRepeatOffenders \textit{repeat offenders}, or
publishers with multiple vetted extensions.
Up to \infringingRepeatOffendersWithPublishedP of these publishers
(\infringingRepeatOffendersWithPublished accounts) have at least one published
infringing extension as of our last date of crawling.
Put another way, \infringingPublishedByRepeatOffendersP of the published
infringing extensions (\infringingPublishedByRepeatOffenders items) come from
repeat offenders.
Another takeaway is that these accounts were not banned from the store after
repeatedly uploading confirmed infringing extensions.
According to the CWS, \textit{``repeated violations''} of the Program Policies will
result in the suspension of the developer account~\cite{chrome-policies-repeat-abuse}.
The existence of known repeat offenders with published extensions is a strong
indication that this policy is not being properly enforced.

\takeaway{%
  We estimate that the CWS fails to detect \publishedExtensionsP of extensions
  which are similar to previously vetted items, suggesting it lacks the ability
  to query extensions based on behavioral similarity.
  As \infringingRepublishedExtensionsP of infringing items are republished
  extensions, using similarity search will greatly improve the vetting process.
  Repeat offenders are also a concern, as \infringingPublishedByRepeatOffendersP
  of published infringing extensions come from accounts with known violations
  that have not been suspended.
}

\subsection{Survival Analysis}
We perform a survival analysis~\cite{survival-analysis-tutorial} on infringing
extensions to assess how rapidly the CWS removes policy-violating content.
To determine how long these extensions remain published in the store, we measure
the \textit{lifetime} or duration in days between the release date of the infringing
extension version and its ``death'' (removal date).
For published infringing extensions---for which we do not observe this event---,
we take the latest date of our crawling as the removal date, and label them as
right-censored observations.

The lifetime distribution for vetted extensions provides some worrying results:
\survivalMoreThanOneYearP of vetted extensions stay in the CWS for more than a
year, with \survivalMoreThanTenYears items even surpassing a decade before
being taken down.
Only \survivalLessThanOneMonthP of these extensions are vetted in less than a
month, and merely \survivalLessThanOneWeekP within a week.
These figures suggest that it takes several months or even years for Google
to remove infringing extensions from the CWS.
During this time, Chromium users can download and install these items without
any notice.

\begin{figure}[t]
  \centering
  \includegraphics[width=0.9\columnwidth]{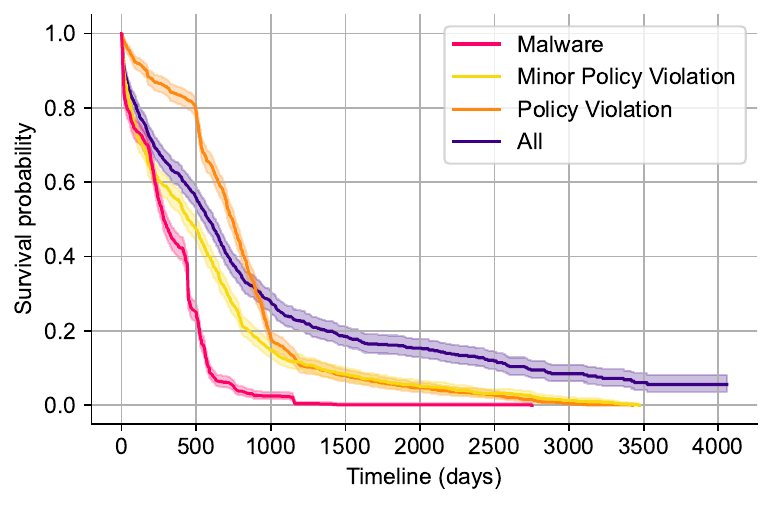}
  \caption{Kaplan--Meier curves of the survival of infringing extensions by vetting label.}
  \label{fig:survival-km}
\end{figure}
To study the likelihood of survival for all infringing items, we use the lifelines
Python library~\cite{lifelines} to plot Kaplan--Meier curves and assist with
related statistical analysis.
When plotting these curves using all \survivalAllSize infringing extensions, the
confidence intervals became too narrow and were not visible in the figure.
Since this makes it difficult to assess the uncertainty around the survival
estimates, we perform several experiments with different random sample sizes to
find the best visualization.
Seeing that the curves have no meaningful visible changes other than the
confidence intervals, we take a random sample of \survivalSampleSize extensions
(per variable or class) to display the Kaplan--Meier curves in a clearer manner.
The curve labeled as ``All'' from Figure~\ref{fig:survival-km} shows the survival
probability of infringing extensions.
The median survival time is \survivalKmMedian days, confirming our earlier
assertion that more than half of vetted extensions remain in the CWS for
more than a year.
However, while the distribution of vetted extensions shows few items with a life
expectancy greater than a decade, the Kaplan--Meier estimate predicts that
\survivalKmProbMoreTenYears of infringing extensions will surpass the same period.
This result makes sense due to the inclusion of published infringing extensions
as censored observations in the calculation of this curve.

We calculate multiple survival curves stratified by vetting label, and look for
differences in the lifetime of vetted extensions based on this variable.
Then, we perform log-rank tests between the three independent classes and obtain
p-values smaller than 0.005 in all cases, thus rejecting the null hypothesis of
no difference between groups.
As such, we conclude that there is likely a significant difference in the lifetime
of vetted extensions based on their vetting label.
Figure~\ref{fig:survival-km} also shows separate Kaplan-Meier curves for these
classes.
We see substantial changes in the median survival time, with extensions flagged as
policy violations having the longest expected lifetime
(\survivalKmPolicyViolationMedian days),
and malware having the shortest (\survivalKmMalwareMedian days).
The longer lifetime of policy violations may be related to the need for user
complaints to initiate the process, although we do not have enough visibility
into the vetting process to confirm this.
Interestingly, policy violations tend to stay on the CWS for longer than
\textit{minor} policy violations (\survivalKmMinorPolicyViolationMedian days),
which is counterintuitive given the label name assigned by Google.

\takeaway{%
  The CWS is excessively slow at removing infringing extensions, typically taking
  months or even years.
  At best, only \survivalLessThanOneMonthP of extensions are vetted in less than
  a month.
  Malware is taken down faster than other types of infringing extensions, with a
  median lifetime of \survivalKmMalwareMedian days or well over half a year.
}

%% file: sections/rq2.tex
\section{Characterizing Infringing Extensions}
\label{sec:rq2}

\input{figures/kinds-clusters-table}

This section examines what kinds of infringing extensions are the most prevalent
in our dataset (RQ2).
For this analysis, we take the infringing clusters obtained in
Section~\ref{sec:rq1} and sort them by descending size, prioritizing those with
the most extensions.
That is, we are interesting in finding the most common groups of behavior across
infringing extensions.
We select the top largest infringing clusters and assign a \textit{kind} or type
of content to each one by manually inspecting a random sample of the extensions
it contains.

Table~\ref{tab:kinds-clusters} shows the ranked list of the top \kindsTopClusters
infringing clusters.
Almost all of these clusters contain NTEs.
To provide a broader picture of what kind of extensions lie in these clusters, we
only display up to \kindsTopNteClusters NTE clusters in
Table~\ref{tab:kinds-clusters}, allowing for other kinds of clusters to appear.
Therefore, rows that are missing from the table, such as cluster \#18,
are omitted because they also contain NTEs.

Top infringing clusters range in size from hundreds to a few thousand
extensions.
The number of unique publishers per cluster is always lower than its size,
confirming that some developers republish the same extension with the same account.
Out of the top \kindsTopClusters infringing clusters, containing
\kindsTotalExtensions extensions, we only find \kindsTotalPublishers publishers.
With some permissible exceptions allowed by the CWS, this is a strong indication
of repetitive content~\cite{chrome-policies-spam}.
NTE clusters have the highest removal rate, with almost no extensions still
published at the end of our crawl.
Interestingly, we find some instances of no consensus over the vetting label
among extensions in the same cluster.
For example, clusters \#2 and \#4 contain almost identical extensions
that have been removed completely from the CWS, yet the reason varies among
Minor Policy Violation, Policy Violation, and Malware.
There is a high variance in the cluster impact, finding some with millions of
affected users.
The most severe case is cluster \#12, which lasted just 2 months in the
store but managed to gather a total of over 150 million users.
In terms of lifetime, top infringing extensions last an average of one year.
Some clusters like \#34 are even fairly new, having been created a year ago
and containing infringing extensions that have not yet been removed from the CWS.

We next characterize the most significant kinds of clusters we find in our
results, providing examples for some of the extensions they contain.

\subsection{New Tab Extensions (NTEs)}
\label{sec:rq2:ntes}
NTEs are, by far, the most repeated kind of extension, accounting for
\kindsTotalTopNtesP of the top infringing clusters, or \kindsTotalTopNtes
extensions alone.
NTEs override the default page that loads whenever a new tab is opened
in the browser, offering more customization options and features.
The common trait of NTEs is that they display a random background image every
time this page loads.
This has contributed to the proliferation of this type of extension in the CWS,
as each NTE focuses on a particular theme or topic upon which the background
images are based.
Topics cover virtually every user interest imaginable, from abstract
(colors, shapes), miscellaneous (cute animals, cars),
places (Tokyo, New York skyline), sports, celebrities (people, music bands)
or works of fiction (movies, video games).

To prevent publishers from flooding the CWS with low-quality, repetitive
extensions, Google released new program policies in June 2021 that prohibited
\textit{``multiple extensions with highly similar functionality,''} specifically
citing \textit{``wallpaper extensions''} as an example~\cite{chrome-ntes-policy-update}.
Consequently, close to all NTEs found in the top infringing clusters have already
been removed.
We hypothesize that this high removal rate, in contrast to non-NTE clusters, is
due to the fact that NTEs are easily findable because they include recurring
keywords such as ``wallpaper'' or ``new tab'' in their names, facilitating the
vetting process.
Nevertheless, we still find \kindsTotalTopPublishedNtes published NTEs in top
infringing clusters, and at least \kindsTotalPublishedNtes in the entire store
(\ie including clusters outside this ranking).

\begin{figure}[t!]
  \centering
  \begin{subfigure}[b]{0.49\columnwidth}
    \centering
    \includegraphics[width=0.9\linewidth]{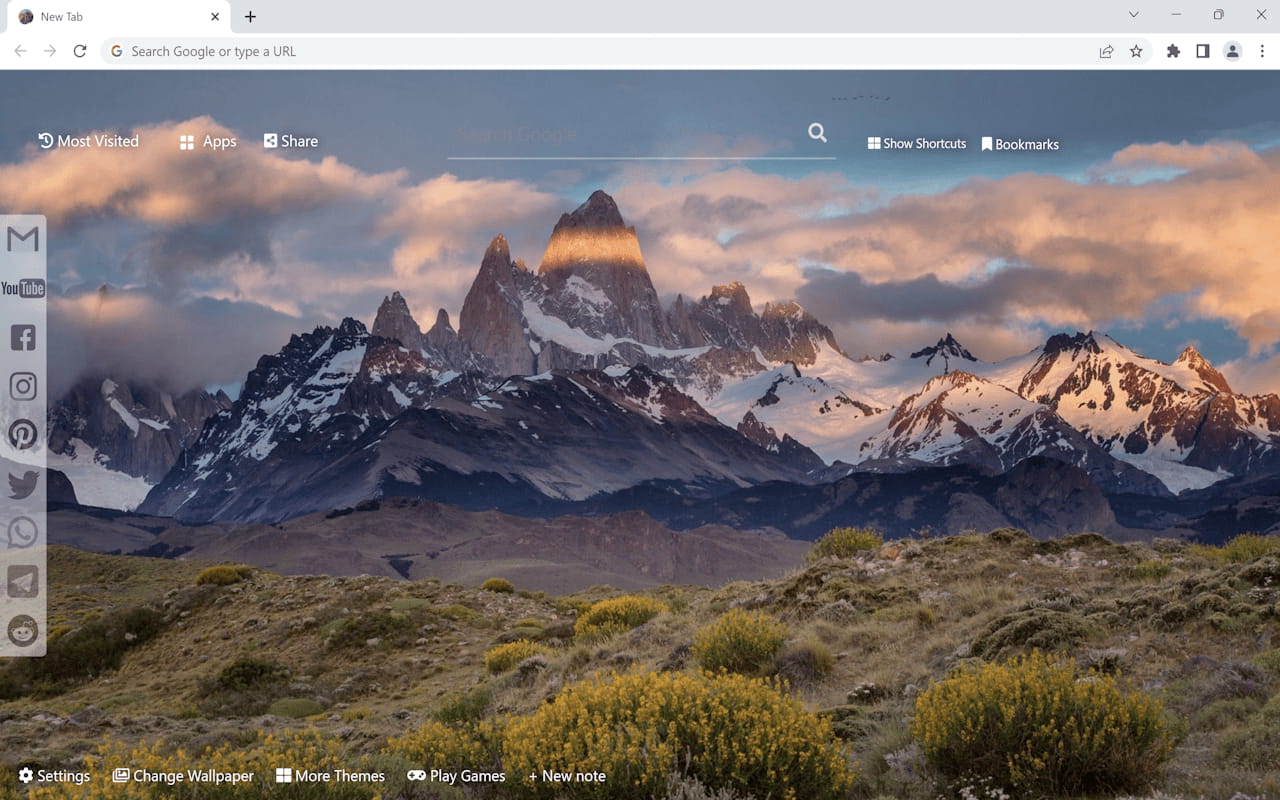}
    \caption{Mountain Wallpaper~\cite{ext-mountain-wallpaper}}
  \end{subfigure}
  \begin{subfigure}[b]{0.49\columnwidth}
    \centering
    \includegraphics[width=0.9\linewidth]{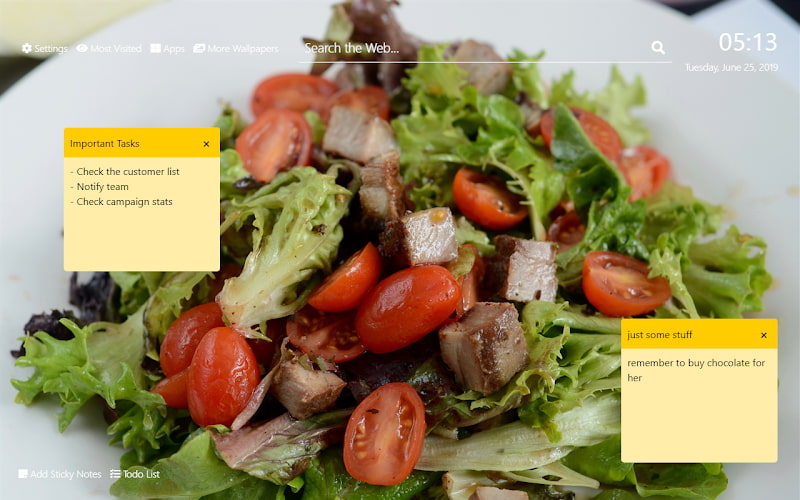}
    \caption{Salads Wallpaper~\cite{ext-salads-wallpaper}}
  \end{subfigure}
  \caption{Sample screenshots of published NTEs.}
  \label{fig:ntes-examples}
\end{figure}
Apart from random background images, NTEs often provide other side features to
stand out from other competing publishers.
As shown in Figure~\ref{fig:ntes-examples}, these features include a navigation
bar with quick links to social media and other popular websites, shortcuts to the
list of installed apps, bookmarks and most visited sites, draggable sticky notes,
realtime weather forecast, and a search bar.
Some extensions go a step beyond and override the default browser search engine,
intercepting all searches typed in the omnibar.

While seemingly innocuous, NTEs may pose a risk to users due to the amount of
personal information they have access to for legitimate purposes.
Overriding the new tab page can be used to track browsing habits, whereas the list
of top visited sites and searches help profile user interests.
For this reason, we believe that clusters \#2 and \#4 contain hundreds
of NTEs labeled as Malware instead of the customary Policy Violation.
However, after manually examining a random subset, we find no meaningful behavioral
differences between malware-labeled extensions and the rest of items in NTE
clusters.
As such, we conclude that CWS's vetting labels are somewhat inconsistent and may
differ between extensions with exactly the same set of behaviors or even codebase.

\subsection{No Content Extensions}
For a extension to be considered eligible for submission to the CWS, it must have
a name, description, version, and an icon declared in its
manifest~\cite{chrome-prepare-your-extension}.
Despite satisfying all the minimum requirements, an extension with only these
attributes would be useless due to its lack of functionality.
Furthermore, according to the Program Policies, it would violate the Minimum
Functionality clause by providing no utility whatsoever to the CWS
catalog~\cite{chrome-policies-minimum-functionality}.

\vspace{1mm}
\noindent\textbf{Test Extensions.}\xspace
Cluster \#33 contains extensions with the bare minimum manifest and an icon,
lacking any HTML, CSS or Java\-Script code.
The items in this cluster date back as early as November 2015, with the most
recent having been creating just a couple of days before the end of our crawl in
\crawlingEnd.
These extensions have names like ``Free Trial Extension!'' or ``Search test,'' and
belong to publishers with dubious email addresses such as ``cwsprodtest3@gmail.com''
or that end at google.com, making the latter corporate Google accounts.
We reported this cluster to the CWS Team asking for a clarification on the purpose
of these extensions and whether they actually were uploaded by their employees.
Unfortunately, we did not receive any response before the submission deadline.

\vspace{1mm}
\noindent\textbf{Boilerplate Extensions.}\xspace
We find another cluster comprised of extensions with no functionality released
between 2018 and 2022.
These extensions have identical dummy background pages and a browser action that
loads the example HTML document from Listing~\ref{lst:extensionizr-html}.
Looking at this document, we notice that the extensions are made using
Extensionizr, a now-defunct web application for generating boilerplate browser
extensions in a few clicks~\cite{extensionizr}.
While the extensions themselves do nothing, their store listings appear to promote
blogs or businesses, with the featured image usually being a screenshot of the
promoted website.
As such, we posit these infringing extensions are just dummy items designed to
take advantage of the CWS to advertise a given site.

\begin{lstlisting}[
  float=t,
  caption={Default browser action document generated by Extensionizr.},
  label={lst:extensionizr-html},
  language=HTML5
]
<!doctype html>
<style type="text/css">
  #mainPopup {
    padding: 10px;
    height: 200px;
    width: 400px;
    font-family: Helvetica, Ubuntu, Arial, sans-serif;
  }
  h1 {
    font-size: 2em;
  }
</style>
<div id="mainPopup">
  <h1>Hello extensionizr!</h1>
  <p>To shut this popup down, edit the manifest file and remove the <@\textcolor{codeDefault}{"default~popup"}@> key. To edit it, just edit ext/browser_action/browser_action.html. The CSS is there, too.</p>
</div>
\end{lstlisting}

\subsection{Spam Extensions}
\label{sec:rq2:spam}
Among the top infringing clusters, we find extensions whose sole purpose is to
promote or spam a particular website or link.
Much like no content extensions, these items provide the user with no apparent
meaningful functionality.
However, unlike the previous group, they do contain source code and assets, albeit
with little effort.
The implementation varies slightly by cluster, with all extensions in a cluster
reusing the same manifest, file structure and much of their source code.

One example is cluster \#16, which consists of extensions only declaring a
browser action in their manifest.
When clicking the extension icon, it loads an HTML document that has a link to
the spammed website.
These extensions rely on popular apps and services to grab the attention of CWS
users, with names such as ``123movie - 100\% WORKING'', ``Free Dogecoin Faucet''
and ``Whatsapp Plus APK''.
Notably, ``GS Auto Clicker:Free Download'' is the only extension labeled as Malware
in this cluster,\footnote{
  \url{https://crxcavator.io/report/chjhnkfpbgcajkfidohljkjjlfcmnahi}
} which differs from the rest only in the link it promotes.
Cluster \#63 is also made up of extensions that have a browser action, but
they take it a step further to give the appearance of having some functionality.
This is achieved by having an HTML form with a few fields that relies on a simple
JavaScript code to produce some---usually meaningless---output.
This form also contains a call-to-action button promoting the spammed website.
For example, the extension ``arsenal vs everton live'' has a football match score
predictor that gives a random pair of numbers every time the form is
submitted~\cite{ext-arsenal-vs-everton}.

\vspace{1mm}
\noindent\textbf{Games.}\xspace
A predominant type of content used to promote websites across top infringing
clusters is games.
These extensions generally follow the same implementation pattern observed in
other spam clusters.
Extensions in clusters \#24 and \#25 implement a browser action with a
``Play'' or ``Play Now'' button that opens the purported URL for the game in a
new tab.
Similarly, clusters \#48 and \#54 have a background page to open this URL
in a new window when the extension icon is clicked.
We also find publishers that make an effort to comply with the Minimum
Functionality clause by embedding the actual game inside the extension instead
of redirecting the user to an external website.
Clusters \#34 and \#65 are instances of this case, containing Unity WebGL
and Adobe Flash games, respectively.
These extensions also have links to websites hosting games, with the distinction
that they can work offline.

\vspace{1mm}
\noindent\textbf{Unreliable Clusters.}\xspace
We find 2 other clusters resembling those containing spam extensions, as
they also have very similar manifests that just declare a browser action.
While these clusters are mostly comprised of extensions with no functionality and
links to external websites, they also contain low-effort extensions that perform
an extremely simple task (\eg a calculator).
Unlike spam extensions, items in unreliable clusters lack a distinct set of API
calls that can be used to group them together, and so they end up in different
clusters than the former.
We acknowledge that this is a limitation of our pipeline as currently designed,
since extensions with short manifests and no API calls are clustered together.
We discuss this point further in Section~\ref{sec:discussion}.

\subsection{Other Extensions}
In addition to the previous groups, we encounter other clusters of extensions
that, while similar, are more difficult to classify as infringing content without
further manual inspection.

\vspace{1mm}
\noindent\textbf{Chrome Apps.}\xspace
Apps are a legacy type of extension that were deprecated in 2020 and are
currently only supported on Chrome\-OS devices~\cite{chrome-apps}.
They can be as simple as a shortcut that opens a URL in a new tab, being
essentially an icon and a manifest that declares the \texttt{app.launch.web\_url}
property.
As such, neither our tool (nor any tool) can determine whether they violate any
policies without analysis of the linked external site.

\vspace{1mm}
\noindent\textbf{Kiosk Apps.}\xspace
Cluster \#19 is formed by kiosk apps generated using the official Chrome
App Builder extension~\cite{ext-chrome-app-builder}.
Akin to Chrome Apps, these extensions just load an external URL, but are intended
to run full-screen on a dedicated kiosk or signage device.

\vspace{1mm}
\noindent\textbf{Repackaged APKs.}\xspace
ChromeOS devices can run Android apps through the App Runtime for Chrome
(ARC)~\cite{arc-announcement}.
With this runtime, developers can bundle Android Application Package (APKs) into
browser extensions and run them on a Chromebook.
Because our pipeline cannot extract features from native APKs, we cannot cluster
these extensions any further.

\vspace{1mm}
\noindent\textbf{Text Replacers.}\xspace
We find a cluster of satirical extensions, often related to politics, with the
sole purpose of replacing a pattern with another word or phrase on all visited
webpages (\eg ``stocks'' to ``stonks'' [\textit{sic}]).
They do so by declaring a content script in their manifest that injects a short
JavaScript file to manipulate the DOM.
We are unsure whether these extensions comply with the Minimum Functionality
and Spam and Abuse policies.
However, we note that a third of this cluster has been taken down from the CWS.

\subsection{Labeling Consistency}
As described in Section~\ref{sec:data-collection}, the CWS assigns a label to
vetted extensions based on the reason for the takedown.
After characterizing the top infringing clusters, we notice some discrepancies
in the labels assigned by Google to these vetted extensions.
This raises the question of whether the CWS vetting labels are indeed related to
the behavior of an extension and if the process is consistent.
To assess this, we identify which are the most common features among vetted
extensions per label and compare the groups.
We take all vetted extensions in our dataset, group them by vetting label, and
count the occurrences of features in each group; that is, we count how many vetted
extensions with a particular label have a given feature.
We initially conducted this experiment including NTEs.
However, doing so added too much noise to our results due to the large number of
vetted NTEs and the seeming inconsistency of their vetting labels, as previously
mentioned in Section~\ref{sec:rq2:ntes}.
As such, we repeated the experiment and put extensions with ``theme'',
``wallpaper'' or ``new tab'' in their name in a separate group.

\input{figures/explainability-vetted-table}
Table~\ref{tab:explainability-vetted} shows the top most occurring features found
in vetted extensions per vetting label.
Overall, there is a considerable overlap between the feature sets of the reported
groups, with 5 features consistently appearing in the top 10 rankings
regardless of the vetting label.
Furthermore, these repeated features are related to the \texttt{storage},
\texttt{runtime} and \texttt{tabs} extension APIs, which are widely used in
published (non-vetted) extensions as well.
Still, we notice some interesting aspects that are worth discussing.

According to Google, one of the key reasons for transitioning to Manifest V3 is its
\textit{``higher security and privacy guarantees''}~\cite{chrome-mv3-transition}.
We see that malware authors are already transitioning to it, with this group having
the highest ratio of extensions using MV3 among the studied groups (40\%).
One of the top API calls used by extensions in the Minor Policy Violation group
is \texttt{browser.\allowbreak{}browserAction.\allowbreak{}onClicked}.
This method is used to add a listener that will run when the extension icon is
clicked, typically to open a new tab or a browser action popup.
This validates the evidence we discuss in Section~\ref{sec:rq2:spam}.
Regarding NTEs, we see that more than half of them declare a URL that will
open in a new tab when the extension is uninstalled, a much higher ratio than for
the other groups.
We also find that the \texttt{topSites} permission is used almost exclusively by
NTEs, making it a good indicator of this group across vetted extensions.

Based on these findings, we conclude that vetting labels are not strictly related to
the behavior of an extension, especially in the case of NTEs, which appear in all
three labels.

\takeaway{%
NTEs are the most common kind of infringing extension, accounting for
\kindsTotalTopNtesP of the top \kindsTopClusters largest clusters.
Extensions with no functionality that clearly do not comply with the store
policies number in the thousands, with hundreds of items still being
published.
CWS vetting labels are somewhat inconsistent, not being strictly related to
the behavior of an extension.
}

%% file: figures/kinds-clusters-table.tex
\begin{table*}[t]
  \centering
  \caption{Top \kindsTopClusters largest infringing clusters. Includes only the first \kindsTopNteClusters NTE clusters, skipping the rest to provide a more representative sample. Footer row has totals from all clusters, including skipped. Highlighted rows contain published extensions. Extension labels are Total (T), Published (P), Policy Violation (PV), Minor Policy Violation (MPV), and Malware (M).}
  \label{tab:kinds-clusters}
  \resizebox{\textwidth}{!}{%
    \small
    \begin{tabular}{rlllrrrrrrrrr}
      \toprule
      \multicolumn{2}{l}{\thead[l]{Cluster}} & \multicolumn{2}{l}{\thead[l]{Example}}    & \multicolumn{5}{l}{\thead[l]{Extensions}}                    & \thead{Publs.}   & \thead{Users}  & \multicolumn{2}{l}{\thead[l]{Lifetime}} \\
      \thead[r]{\#} & \thead[l]{Kind}        & \thead[l]{Extension ID} & \thead[l]{Name} & \thead{T} & \thead{P} & \thead{PV} & \thead{MPV} & \thead{M} & \thead[r]{Count} & \thead[r]{Sum} & \thead[r]{Avg} & \thead[r]{Max}         \\
      \midrule
      1 & NTEs & mngkjbegjgngjmbpojnmkngelecdodfm & Fruits Basket New Tab... & 3.6k & \textcolor{gray}{0} & 1.5k & 2.1k & 1 & 520 & 153k & 1y & 4y \\
      2 & NTEs & cpcnejdpjegmjjljgfjklloeihdkdgbp & Island Wallpapers The... & 2.5k & \textcolor{gray}{0} & 829 & 899 & 822 & 258 & 181k & 1y & 1y \\
      3 & NTEs & cbhjljcgicdehmecljnoggonepjpflij & Tuzki New Tab Page HD... & 2.5k & \textcolor{gray}{0} & 17 & 2.4k & \textcolor{gray}{0} & 253 & 42k & 1y & 1y \\
      4 & NTEs & nidmpdeffdnhnccmmkfaagfoefhajmfl & Lexus Wallpapers HD T... & 1.8k & \textcolor{gray}{0} & 698 & 735 & 255 & 134 & 82k & 1y & 1y \\
\rowcolor{cyan!5}
      5 & NTEs & meppkkplebidmpambckjodfhgheipfdp & Superhero.io HD Wallp... & 1.7k & 3 & 69 & 1.6k & \textcolor{gray}{0} & 181 & 121k & 1y & 3y \\
      6 & NTEs & fhidgmalgieecohkfdmehekjejlkiblo & 4K Wallpaper HD Custo... & 1.6k & \textcolor{gray}{0} & 33 & 573 & 9 & 145 & 18M & 6m & 1y \\
      7 & NTEs & gdjpneilpaakfdkbofbdcoggimplfjek & Dragon Ball Super Bro... & 1.5k & \textcolor{gray}{0} & 240 & 1.3k & \textcolor{gray}{0} & 132 & 15k & 2y & 3y \\
      8 & NTEs & gfdbfnafbpicmggajcjkehmdcnoadigh & Sports car HD wallpap... & 1.5k & \textcolor{gray}{0} & 149 & 1.4k & \textcolor{gray}{0} & 156 & 33k & 5m & 6m \\
      9 & NTEs & lohlomonaokoijibhiaofphjkhpdmhhd & Evan Fournier Themes ... & 1.5k & \textcolor{gray}{0} & 168 & 1.0k & \textcolor{gray}{0} & 144 & 78k & 1y & 4y \\
\rowcolor{cyan!5}
      10 & Unreliable & dmhapbkbkeopdeapenongacbpajdfljg & Calc LS & 1.3k & 53 & 49 & 1.1k & 2 & 292 & 1M & 4m & 10y \\
      11 & NTEs & jkocidindgkhipkkiaffdmoonpdpjkoo & Bridgerton Wallpapers... & 1.3k & \textcolor{gray}{0} & 2 & 783 & \textcolor{gray}{0} & 80 & 174k & 2m & 4m \\
      12 & NTEs & enihnbifjcjijpacpllggccehdlcomco & My New Tab extension ... & 1.2k & \textcolor{gray}{0} & \textcolor{gray}{0} & 1.2k & 2 & 91 & 154M & 1m & 2m \\
\rowcolor{cyan!5}
      13 & NTEs & emkadehobopegnnodjgmifldohhhehae & Cityscape - City Wall... & 1.2k & 1 & 7 & 1.1k & \textcolor{gray}{0} & 79 & 5M & 8m & 1y \\
      14 & NTEs & ngeclbmgabdohlkcpdjbgpoplghdhgbn & Will Smith Popular St... & 1.2k & \textcolor{gray}{0} & 669 & 491 & \textcolor{gray}{0} & 199 & 46k & 2y & 3y \\
\rowcolor{cyan!5}
      15 & NTEs & ckkplkjpdkpbnlcbbhldphbfmgcdfgcn & Smite Wallpapers New ... & 1.1k & 1 & 199 & 923 & \textcolor{gray}{0} & 100 & 123k & 2y & 3y \\
\rowcolor{cyan!5}
      16 & Spam & fpfieneffjbipoalhjkbjdeophjfgmhj & WhatsApp Group Links & 1.0k & 67 & 59 & 740 & 1 & 300 & 545k & 8m & 3y \\
      17 & NTEs & jafpgkdldemnlbmmcehiecehpigjploo & Rwanda New Tab Rwanda... & 905 & \textcolor{gray}{0} & 297 & 585 & 2 & 76 & 54k & 1y & 1y \\
\rowcolor{cyan!5}
      19 & Kiosk & nllpcefijadedeobnimfhahfpacimcgk & Endless Runner & 878 & 717 & 25 & 77 & \textcolor{gray}{0} & 612 & 11k & 5y & 9y \\
      24 & Games & afbcbdeiiofjclcghcjiackbcnebhlhi & Umbrella Down & 508 & \textcolor{gray}{0} & 3 & 505 & \textcolor{gray}{0} & 44 & 10k & 1y & 2y \\
      25 & Games & gaiajadlfkefhennokenpgbflgmfelej & WorldCraft: 3D Build ... & 501 & \textcolor{gray}{0} & 4 & 493 & \textcolor{gray}{0} & 22 & 29k & 1m & 2y \\
\rowcolor{cyan!5}
      29 & Apps & cgpdaffgnmiapcmnmjlccgchnlnddhch & Suecalandia & 423 & 370 & 1 & 31 & \textcolor{gray}{0} & 377 & 2M & 8y & 11y \\
\rowcolor{cyan!5}
      33 & Test & nlmjjcpjklaapbblhokbnfbjahjodfkh & Erase From View For T... & 372 & 67 & 7 & 35 & 6 & 270 & 1.2k & 1m & 7y \\
\rowcolor{cyan!5}
      34 & Games & mbmibbpjgfmhodpjhgcollomgggpkpbn & Balls Avoid Arcade Ga... & 316 & 300 & 10 & 6 & \textcolor{gray}{0} & 45 & 1M & 1y & 1y \\
\rowcolor{cyan!5}
      35 & Replacer & lijiloghjiehhmggfgloglklmbnoenen & Impeachment Pie & 314 & 198 & 11 & 95 & \textcolor{gray}{0} & 300 & 11k & 5y & 8y \\
\rowcolor{cyan!5}
      47 & Unreliable & niiicdjkhpheppjjaankjoegbbejdmho & Magento Development C... & 266 & 38 & 11 & 197 & \textcolor{gray}{0} & 237 & 64k & 3y & 5y \\
      48 & Games & himbonlbdfgjpnfdabgkomgghffbfhph & Tarot Kariyer Falı Ba... & 264 & \textcolor{gray}{0} & 1 & 254 & 2 & 80 & 13k & 5m & 1y \\
\rowcolor{cyan!5}
      49 & APKs & mdmkciaalenjmmannlibndcgjjiibdng & ce & 263 & 242 & 1 & 4 & \textcolor{gray}{0} & 196 & 72k & 7y & 9y \\
      54 & Games & gnpikgcjfmlcjpcnokpiindllmdmadem & Digimon Battle Spirit... & 229 & \textcolor{gray}{0} & 1 & 227 & 1 & 22 & 23k & 7m & 3y \\
\rowcolor{cyan!5}
      60 & Boilerplate & aemnapldoeccnbmfmifibgdjimkdgnig & League of Legends Acc... & 210 & 76 & 16 & 108 & \textcolor{gray}{0} & 160 & 28k & 4y & 5y \\
\rowcolor{cyan!5}
      63 & Spam & bbcfdgkbbllgjidioakddokoakeecloj & Group Buy Seo Tools & 206 & 189 & \textcolor{gray}{0} & 11 & \textcolor{gray}{0} & 125 & 1.6k & 11m & 2y \\
      65 & Games & kjdcdphljphkdkipfhijimoedhpeheop & Epic Charlie & 201 & \textcolor{gray}{0} & 2 & 199 & \textcolor{gray}{0} & 14 & 4.3k & 1y & 3y \\
      \midrule
      &  &  &  & 44k & 2.4k & 6.4k & 29k & 1.6k & 6.2k & 218M & 1y & 11y \\
      \bottomrule
    \end{tabular}
  }
\end{table*}

%% file: figures/explainability-vetted-table.tex
\begin{table}[t]
  \centering
  \caption{Top most occurring features in vetted extensions grouped by label. NTEs reported separately, highlighted rows show features in any top 5.}
  \label{tab:explainability-vetted}
  \resizebox{\columnwidth}{!}{%
    \setlength{\tabcolsep}{0.5em}
    \small
    \begin{tabular}{lp{7pt}rp{7pt}rp{7pt}rp{7pt}r}
      \toprule
      \thead[l]{Feature} & \multicolumn{2}{r}{\thead[r]{M}} & \multicolumn{2}{r}{\thead[r]{PV}} & \multicolumn{2}{r}{\thead[r]{MPV}} & \multicolumn{2}{r}{\thead[r]{NTE}} \\
      \midrule
\rowcolor{cyan!5}
      MV2 & {\scriptsize\textcolor{yellow!90!black}{\textbf{\#1}}} & 60\% & {\scriptsize\textcolor{yellow!90!black}{\textbf{\#1}}} & 86\% & {\scriptsize\textcolor{yellow!90!black}{\textbf{\#1}}} & 93\% & {\scriptsize\textcolor{brown}{\textbf{\#3}}} & 88\% \\
\rowcolor{cyan!5}
      storage & {\scriptsize\textcolor{gray}{\textbf{\#2}}} & 52\% & {\scriptsize\textcolor{gray}{\textbf{\#2}}} & 45\% & {\scriptsize\textcolor{cyan!90!black}{\textbf{\#4}}} & 30\% & {\scriptsize\textcolor{cyan!30!black}{\textbf{\#8}}} & 46\% \\
\rowcolor{cyan!5}
      browser.tabs.create & {\scriptsize\textcolor{brown}{\textbf{\#3}}} & 51\% & {\scriptsize\textcolor{cyan!60!black}{\textbf{\#5}}} & 34\% & {\scriptsize\textcolor{brown}{\textbf{\#3}}} & 30\% & {\scriptsize\textcolor{gray}{\textbf{\#2}}} & 93\% \\
\rowcolor{cyan!5}
      browser.runtime.onInstalled & {\scriptsize\textcolor{cyan!90!black}{\textbf{\#4}}} & 49\% & {\scriptsize\textcolor{cyan!30!black}{\textbf{\#8}}} & 22\% & {\scriptsize\textcolor{cyan!30!black}{\textbf{\#8}}} & 19\% & {\scriptsize\textcolor{cyan!90!black}{\textbf{\#4}}} & 87\% \\
\rowcolor{cyan!5}
      MV3 & {\scriptsize\textcolor{cyan!60!black}{\textbf{\#5}}} & 40\% & {\scriptsize\textcolor{black}{\textbf{\#21}}} & 14\% & {\scriptsize\textcolor{black}{\textbf{\#31}}} & 7\% & {\scriptsize\textcolor{black}{\textbf{\#57}}} & 12\% \\
\rowcolor{cyan!5}
      browser.runtime.onMessage & {\scriptsize\textcolor{cyan!50!black}{\textbf{\#6}}} & 40\% & {\scriptsize\textcolor{cyan!90!black}{\textbf{\#4}}} & 35\% & {\scriptsize\textcolor{cyan!60!black}{\textbf{\#5}}} & 21\% & {\scriptsize\textcolor{cyan!10!black}{\textbf{\#10}}} & 43\% \\
      browser.runtime.id & {\scriptsize\textcolor{cyan!40!black}{\textbf{\#7}}} & 37\% & {\scriptsize\textcolor{black}{\textbf{\#30}}} & 11\% & {\scriptsize\textcolor{black}{\textbf{\#29}}} & 7\% & {\scriptsize\textcolor{cyan!40!black}{\textbf{\#7}}} & 47\% \\
      browser.storage.local.get & {\scriptsize\textcolor{cyan!30!black}{\textbf{\#8}}} & 37\% & {\scriptsize\textcolor{black}{\textbf{\#12}}} & 20\% & {\scriptsize\textcolor{black}{\textbf{\#21}}} & 10\% & {\scriptsize\textcolor{black}{\textbf{\#23}}} & 22\% \\
\rowcolor{cyan!5}
      tabs & {\scriptsize\textcolor{cyan!20!black}{\textbf{\#9}}} & 36\% & {\scriptsize\textcolor{brown}{\textbf{\#3}}} & 41\% & {\scriptsize\textcolor{gray}{\textbf{\#2}}} & 42\% & {\scriptsize\textcolor{black}{\textbf{\#20}}} & 23\% \\
      browser.storage.local.set & {\scriptsize\textcolor{cyan!10!black}{\textbf{\#10}}} & 35\% & {\scriptsize\textcolor{black}{\textbf{\#13}}} & 19\% & {\scriptsize\textcolor{black}{\textbf{\#22}}} & 10\% & {\scriptsize\textcolor{black}{\textbf{\#22}}} & 22\% \\
      browser.tabs.query & {\scriptsize\textcolor{black}{\textbf{\#13}}} & 31\% & {\scriptsize\textcolor{cyan!50!black}{\textbf{\#6}}} & 29\% & {\scriptsize\textcolor{cyan!20!black}{\textbf{\#9}}} & 18\% & {\scriptsize\textcolor{black}{\textbf{\#11}}} & 39\% \\
      browser.runtime.sendMessage & {\scriptsize\textcolor{black}{\textbf{\#12}}} & 32\% & {\scriptsize\textcolor{cyan!40!black}{\textbf{\#7}}} & 27\% & {\scriptsize\textcolor{cyan!10!black}{\textbf{\#10}}} & 14\% & {\scriptsize\textcolor{black}{\textbf{\#13}}} & 36\% \\
      activeTab & {\scriptsize\textcolor{black}{\textbf{\#53}}} & 7\% & {\scriptsize\textcolor{cyan!20!black}{\textbf{\#9}}} & 22\% & {\scriptsize\textcolor{cyan!40!black}{\textbf{\#7}}} & 21\% & {\scriptsize\textcolor{black}{\textbf{\#38}}} & 17\% \\
      navigator.userAgent & {\scriptsize\textcolor{black}{\textbf{\#11}}} & 32\% & {\scriptsize\textcolor{cyan!10!black}{\textbf{\#10}}} & 21\% & {\scriptsize\textcolor{black}{\textbf{\#13}}} & 11\% & {\scriptsize\textcolor{black}{\textbf{\#59}}} & 10\% \\
\rowcolor{cyan!5}
      browser.browserAction.onClicked & {\scriptsize\textcolor{black}{\textbf{\#22}}} & 18\% & {\scriptsize\textcolor{black}{\textbf{\#15}}} & 16\% & {\scriptsize\textcolor{cyan!50!black}{\textbf{\#6}}} & 21\% & {\scriptsize\textcolor{cyan!60!black}{\textbf{\#5}}} & 76\% \\
\rowcolor{cyan!5}
      chrome\_url\_overrides.newtab & {\scriptsize\textcolor{black}{\textbf{\#34}}} & 11\% & {\scriptsize\textcolor{black}{\textbf{\#43}}} & 7\% & {\scriptsize\textcolor{black}{\textbf{\#28}}} & 7\% & {\scriptsize\textcolor{yellow!90!black}{\textbf{\#1}}} & 99\% \\
      browser.runtime.setUninstallURL & {\scriptsize\textcolor{black}{\textbf{\#14}}} & 26\% & {\scriptsize\textcolor{black}{\textbf{\#36}}} & 9\% & {\scriptsize\textcolor{black}{\textbf{\#35}}} & 6\% & {\scriptsize\textcolor{cyan!50!black}{\textbf{\#6}}} & 57\% \\
      topSites & {\scriptsize\textcolor{black}{\textbf{\#262}}} & 1\% & {\scriptsize\textcolor{black}{\textbf{\#129}}} & 2\% & {\scriptsize\textcolor{black}{\textbf{\#66}}} & 3\% & {\scriptsize\textcolor{cyan!20!black}{\textbf{\#9}}} & 45\% \\
      \bottomrule
    \end{tabular}%
  }
  \end{table}

%% file: sections/rq3.tex
\section{Characterizing CWS Malware}
\label{sec:rq3}

Out of all vetted extensions, those labeled as malware are of significant
importance.
In this section, we study which malware families are published in the CWS and use
\textsc{SimExt} to cluster them by their behavior (RQ3).
We conclude with an analysis of malicious extensions with a high number of
installs that are not present in VirusTotal.
We note that we do not claim to study the entire browser extension malware
ecosystem, as some families may be distributed outside of the CWS by other means,
such as sideloading~\cite{report-rilide, report-parasitesnatcher,
report-shampoo}.

\subsection{Malware Detection}
As mentioned throughout this paper, the CWS assigns its own labels, including a
generic one for malware, to vetted extensions as part of the vetting process.
As such, we ask ourselves how aware the threat intelligence community is about
these extensions that the CWS considers malicious.
To answer this question, we take all vetted extensions in our dataset that are
labeled as Malware, compute the SHA-256 hashes of their CRX packages, and query
VirusTotal to obtain labels from third-party security vendors.
Based on these results, we group our subset of malware extensions into three
categories:
$(i)$ \textit{Not Found} if the sample has not been uploaded to VirusTotal,
$(ii)$ \textit{Clean} for samples with zero detections, and
$(iii)$ \textit{Malicious} if it is flagged as malicious or suspicious by at
least one engine.

Of the \malwareExtensions extensions labeled by the CWS as Malware, a remarkable
\malwareNotFoundP of them (\malwareNotFound extensions) were not previously seen
by VirusTotal;
\malwareCleanP of malware extensions fall into the Clean category; and merely
\malwareMalicious extensions (\malwareMaliciousP) are detected as Malicious.
Having such a significantly high number of Not Found and Clean items for extensions
that the CWS itself reports as malware raises some concerns.
One implication is that most malware families that target the CWS are unknown to
the threat intelligence community, possibly because these samples are just removed
from the CWS (and from infected victims) but they are not shared with the community.

\subsection{Malware Families}
We group the few known malicious samples that we found by their most popular threat
classification, or \textit{family}, as suggested by VirusTotal.
For each group, we count the number of extensions, measure the impact by the number
of user installs, the coverage of the detection engines by the ratio of detections,
and the lifetime of a family by the release dates of the first and last seen
extension versions.

\input{figures/malware-families-table}
Table~\ref{tab:malware-families} shows the aforementioned groups of extensions
that are known to VirusTotal per family.
Almost half of the malicious extensions have no suggested threat classification
label, closely followed by a generic ``Trojan'' group.
Other labels such as ``BrowExt'', ``BroExtension'' or ``ChromeX'' seem to only
indicate that the malware is a browser extension, without adding any further
information about the malware family or behavior.
The only three extensions with more informative classifications are
``SaveProtect VPN'',\footnote{
  \url{https://crxcavator.io/report/dodnpoijjkmcmlhlelmggejhfocfjgfc}
} which is labeled as Adware;
``Fea KeyLogger'',\footnote{
  \url{https://crxcavator.io/report/fgkghpghjcbfcflhoklkcincndlpobja}
} which advertises itself as a keylogger;
and ``Free YouTube Subscribers Generator'',\footnote{
  \url{https://crxcavator.io/report/fdfchfidjajpidpjilnlboncflgnjdda}
} which has a link to an external website instead of having any JavaScript code.
From these findings, we conclude that vendor labels for malicious browser
extensions are extremely poor, and in most cases non-existent.

Grouping only these known malicious extensions by behavior using \textsc{SimExt},
we end up with \malwareMaliciousClusters different similarity clusters and
\malwareMaliciousOutliers extensions that are classified as outliers.
The resulting clusters have extensions that fall into the following types:
$(i)$ media file downloaders,
$(ii)$ cursor icon customizers,
$(iii)$ volume boosters,
$(iv)$ reader mode extensions,
and $(v)$ ad blockers.
Appendix~\ref{sec:appendix-malware} lists these extensions for more information.

\subsection{Case Studies}
Since almost all malware extensions flagged by the CWS are unknown to detection
engines, we manually inspect those not present in VirusTotal and with at
least 100k installs at the time of removal.
Of the resulting 90 items that meet these criteria, we could not find a clear
motivation for labeling 29 of them as malware.

A large group of the remaining extensions contain extensions with embedded
tracking capabilities or that load remotely hosted code.
One interesting group is formed by 12 extensions that use Google Tag Manager
to download and execute an obfuscated script\footnote{
  \url{https://storage.googleapis.com/g1analytics/cloud_new_noab-obf.js}
} from a Google Cloud Storage bucket.
When unpacked, this script sends an HTTP request to a remote endpoint to get the
country of the user and starts exfiltrating all visited URLs in real time if the
location matches the United States.
To avoid detection, the payload of these requests is encoded as a binary blob.
Some extensions that request the \texttt{webRequestBlocking} permission load an
additional remote script,\footnote{
  \url{https://storage.googleapis.com/analytics-cloud/js_analytics_protected.js}
} which intercepts all search queries from a predefined list of lesser known search
engines and redirects them to another domain.

We also find 4 trojanized extensions that run code in Facebook to obtain
authentication tokens (such as \texttt{fb\_dtsg}).
The intercepted tokens are then used to make
requests to the GraphQL endpoint in the background without the user's awareness.
Once authenticated, the user is joined to a Facebook group that varies depending
on the extension, and the device's access token is posted there.
Along with the token, these extensions also exfiltrate the number of business
accounts associated with that credential, presumably to triage victims.

Appendix~\ref{sec:appendix-case-studies} lists the extensions discussed
above.

\takeaway{%
Only \malwareMaliciousP of extensions labeled as malware by the CWS are considered
malicious by third-party security vendors integrated on VirusTotal.
Furthermore, these vendors rarely assign a malware family to these samples, and
when they do, they use generic labels that provide little information.
}

%% file: figures/malware-families-table.tex
\begin{table}[t]
  \centering
  \caption{Families of malware extensions detected by engines in VirusTotal.}
  \label{tab:malware-families}
  \resizebox{\columnwidth}{!}{%
    \setlength{\tabcolsep}{0.5em}
    \small
    \begin{tabular}{lrrrrlr}
      \toprule
      \multirowthead{2}{} & \thead[l]{Exts.} & \multicolumn{2}{l}{\thead[l]{Detections}} & \thead[l]{Users} & \multicolumn{2}{l}{\thead[l]{Lifetime}} \\
                          & \thead[r]{Count} & \thead[r]{Max} & \thead[r]{Avg}           & \thead[r]{Sum}   & \thead[l]{From} & \thead[r]{To}         \\
      \midrule
      \textcolor{gray}{N/A} & 39 & 39\% & 10\% & 9M & 2018-01-31 & +5y \\
      trojan. & 28 & 17\% & 13\% & 59M & 2020-05-08 & +2y \\
      browext & 6 & 17\% & 13\% & 9M & 2022-02-23 & +1y \\
      trojan.browext/chromex & 4 & 21\% & 21\% & 365k & 2022-08-26 & +1m \\
      trojan.chromex & 3 & 32\% & 25\% & 2M & 2022-01-19 & +10m \\
      adware.broextension & 1 & 3\% & 3\% & 700k & 2023-10-23 & +1d \\
      pua.keylogger/chromelogger & 1 & 33\% & 33\% & 50k & 2020-03-01 & +1d \\
      trojan.freesub/chromex & 1 & 37\% & 37\% & 8.0k & 2020-11-23 & +1d \\
    \bottomrule
    \end{tabular}
  }
\end{table}

%% file: sections/discussion.tex
\section{Discussion}
\label{sec:discussion}

This section discusses the key findings of our work and provide recommendations
that could help improve the CWS vetting process.
We also discuss the limitations of our analysis, particularly of \textsc{SimExt},
and describe future work to address them.

\vspace{1mm}
\noindent\textbf{Detection of Infringing Extensions.}\xspace
The findings outlined in Section~\ref{sec:rq1} support the assumption that the
current CWS vetting process lacks scalable tooling for effectively finding or
clustering similar extensions.
Given the estimate that \infringingRepublishedExtensionsP of infringing extensions
found by our tool are republished extensions (\ie items that the store has seen
and taken down before), we recommend using not only blocklists but also
behavioral features to flag similar extensions, as this can contribute to
improve the vetting process.
We believe that similarity search tools such as \textsc{SimExt} can be a valuable
complement to the human factor, helping to find infringing content faster and more
accurately.

\vspace{1mm}
\noindent\textbf{Repeat Offenders.}\xspace
We find evidence that the CWS does not ban publishers with multiple vetted
extensions.
Since \infringingPublishedByRepeatOffendersP of the published infringing extensions
found by our tool come from repeat offenders, we believe the CWS should enforce
their own Repeat Abuse policy~\cite{chrome-policies-repeat-abuse} and suspend the
accounts of repeat offenders.

\vspace{1mm}
\noindent\textbf{Lifetime of Infringing Extensions.}\xspace
Infringing content stays for too long in the store, with \survivalMoreThanOneYearP
of vetted extensions remaining published for more than a year before they are
taken down.
Even for malware extensions, the median removal time is longer than 9 months
at best.
We consider these to be unacceptably high and believe that the CWS needs to focus
its efforts on significantly reducing them.
The addition of automated tools for finding items similar to previously vetted
extensions should also help reduce these lifetimes.

\vspace{1mm}
\noindent\textbf{Unpublished Infringing Extensions.}\xspace
As presented in Section~\ref{sec:rq1}, we find that some publishers of infringing
extensions voluntarily remove them before receiving a takedown from Google.
Given that repeat offenders are fairly common (and thus there is little repercussion
for having multiple vetted extensions), we do not believe that these developers are
acting like this to protect their accounts.
Extensions labeled by Google as malware are remotely and automatically uninstalled
from users browsers~\cite{extension-safety-hub}.
However, we find that extensions that are unpublished before being detected as
malware are not subject to this process, leaving users compromised.
Thus, we recommend that the CWS also assigns vetting labels to recently unpublished
extensions as well as to vetted ones.

\vspace{1mm}
\noindent\textbf{Vetting Labels.}\xspace
Providing labels for vetted content is a valuable transparency feature of the CWS
vetting process.
However, the current labeling scheme admits multiple improvements.
One would be to move towards finer-grained labels that are more specific about
the reasons for taking down the item, as these labels could better inform ecosystem
studies such as ours.
As mentioned in Section~\ref{sec:rq2:ntes}, more consistency when assigning
labels to vetted extensions will also help increase confidence in the vetting
process.

\vspace{1mm}
\noindent\textbf{Malware Labels.}\xspace
Given that only \malwareMaliciousP of extensions labeled as malware by the CWS
are detected by security vendors in VirusTotal, we strongly recommend that the
threat intelligence community improves malware detectors for browser extensions.
More fine-grained family labels will greatly assist in classifying and tracking
trends in the browser extension malware ecosystem.

\subsection{Limitations and Future Work}
In Section~\ref{sec:rq2:spam}, we discuss how our pipeline clusters together
extensions with very few features, to the point that items with no logged API
calls and different behaviors are difficult to distinguish from one another.
Ideally, these unreliable clusters should be split further into as many
sub-clusters as unique sets of behaviors.
As an improvement, we propose expanding the API calls extracted by our static
and dynamic analyses to include other frequently used objects, such as
\texttt{document} and \texttt{jQuery}.
This way, extensions with few or no calls to Extension APIs or inside the
\texttt{navigator} object will be better clustered due to the use of more robust
embeddings.

%% file: sections/related-work.tex
\section{Related Work}
\label{sec:related-work}

Previous research has crawled the CWS for various purposes related to the analysis
of browser extensions~\cite{hulk, exray, i-spy-with-my-little-eye, mystique,
youve-changed, doublex, fingerprinting-in-style, dangers-of-human-touch,
hardening-browser-extensions, no-signal-left-to-chance, fakex}.

\vspace{1mm}
\noindent\textbf{Extension Analysis.}\xspace
Pantelaios \etal statically extracted API calls to cluster extension version
deltas to detect malicious updates~\cite{youve-changed}.
Fass \etal used static analysis to find suspicious data flows in vulnerable
extensions~\cite{doublex}.
Kapravelos \etal proposed a dynamic analysis architecture for identifying
malicious behavior in extensions using on-the-fly generated pages~\cite{hulk}.
Jagpal \etal used both static and dynamic analysis to capture behavioral signals
from browser extensions~\cite{trends-and-lessons}, followed by Aggarwal \etal a few
years later~\cite{i-spy-with-my-little-eye}.
Picazo-Sanchez \etal used static, manual and dynamic analysis to mark malicious
extensions~\cite{no-signal-left-to-chance}.
Eriksson \etal combined both static and dynamic analysis to discover code
vulnerabilities, uncovering the existence of NTEs that stole browsing
traffic~\cite{hardening-browser-extensions}.
Pantelaios and Kapravelos used dynamic execution to validate automated conversions
of extensions from MV2 to MV3~\cite{mv3-unveiled}.
Unlike our work, previous dynamic analysis efforts have relied on running the
extensions on an actual web browser through manual interaction or using tools
like Selenium, Puppeteer and Playwright.
Instead, we created a mocked V8 sandbox for faster code execution at scale.

\vspace{1mm}
\noindent\textbf{Defining Potentially Harmful Extensions.}\xspace
Somé was the first to propose a taxonomy of \textit{sensitive APIs} to define
extensions that pose a security or privacy risk to the user~\cite{empoweb}.
Hsu \etal introduced the concept of \textit{Security-Noteworthy Extensions}
to extend this set to cover malware, vulnerable, and policy-violating
extensions~\cite{what-is-in-the-cws}.
In our work, we focus on evaluating the CWS vetting process by using its own
vetting labels, rather than seeking to define what potentially harmful
extensions are.

\vspace{1mm}
\noindent\textbf{Finding Malicious Extensions with Machine Learning.}\xspace
Jagpal \etal used machine learning to flag malicious extensions based on behavioral
signals, having to train a proprietary model daily to account for newly vetted
extensions~\cite{trends-and-lessons}.
Aggarwal \etal improved upon this by feeding sequences of API calls to their own
Recurrent Neural Network (RNN)~\cite{i-spy-with-my-little-eye}.
Pantelaios \etal used DBSCAN to cluster custom-made API sequences referred to as
\textit{seeds}~\cite{youve-changed}.
Similarly, Picazo-Sanchez \etal used time series analysis and machine learning to
cluster malicious extensions based on their download patterns~\cite{no-signal-left-to-chance}.
In contrast to previous work, our approach does not try to classify extensions as
benign or malicious, but to cluster similar extensions together.
By using NLP and ZSL, we avoid the limitation of having to retrain our model when
a new malicious behavior is discovered.

To the best of our knowledge, no prior work has developed a comprehensive
methodology for measuring similarity between extension based on static features
and dynamic behavior, nor used vector embeddings to cluster browser extensions.

%% file: sections/conclusion.tex
\section{Conclusion}
\label{sec:conclusion}

This paper has presented a comprehensive study of the CWS vetting process and the
prevalence of infringing content.
To assist in this analysis, we developed \textsc{SimExt}, a novel methodology for
measuring similarity among browser extensions.
Our tool has proven to be instrumental to find infringing content that otherwise
went unnoticed.
We also believe that \textsc{SimExt} may be valuable in other application domains.
Informed by our findings, we present several recommendations that could contribute
to improving the vetting process and the analysis of malicious and infringing
content.

%% file: sections/appendix-serialization.tex
\section{Serialization Example}
\label{sec:appendix-serialization}

\begin{figure}[ht!]
  \centering
  \includegraphics[width=0.9\columnwidth]{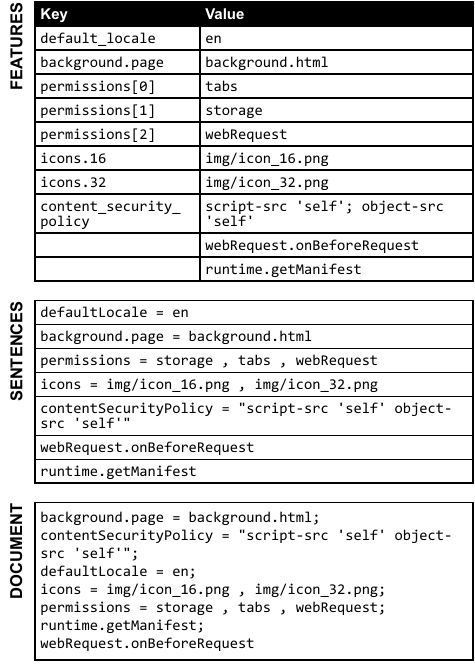}
  \caption{Feature serialization example.}
  \label{fig:serialization-example}
\end{figure}

Figure~\ref{fig:serialization-example} provides an example of the feature
serialization pipeline we describe in Section~\ref{sec:methodology:vectorization}.
The process takes as input the manifest key-value pairs of the extension and the
API calls extracted by the static and dynamic analyzers.
These items are converted to sentences that can then be merged into a single
document.

%% file: sections/appendix-malware.tex
\section{Malicious Clustered Extensions}
\label{sec:appendix-malware}

Table~\ref{tab:malware-clusters} lists extensions detected as Malicious by
third-party security vendors that belong to a similarity cluster obtained
using \textsc{SimExt}.

\begin{table}[ht!]
  \centering
  \caption{Known malicious extensions that belong to a similarity cluster.}
  \label{tab:malware-clusters}
  \resizebox{\columnwidth}{!}{%
    \setlength{\tabcolsep}{0.5em}
    \small
    \begin{tabular}{ll}
      \toprule
      \thead[l]{Extension}             & \thead[l]{Name}               \\
      \midrule
      \rowcolor{black!10} \multicolumn{2}{l}{Media downloaders}        \\
      pocmgnhmjgjghodelfkhbjaoidmbadpo & Spotify™ \& Deezer™ Music ... \\
      nhemnekeahfdfemcchogmiinhkpdbedp & Spotify™ \& Deezer™ Music ... \\
      hpbohmeoofibpbiiklpofdfehodejbmk & VLC Video Downloader          \\
      dhnkeanajeheaifbmgalejfoebggoggk & VLC Video Downloader          \\
      nadenkhojomjfdcppbhhncbfakfjiabp & Base Image Downloader         \\
      okclicinnbnfkgchommiamjnkjcibfid & Tap Image Downloader          \\
      \rowcolor{black!10} \multicolumn{2}{l}{Cursor customizers}       \\
      magnkhldhhgdlhikeighmhlhonpmlolk & Craft Cursors                 \\
      pbdpfhmbdldfoioggnphkiocpidecmbp & Clickish fun cursors          \\
      hdgdghnfcappcodemanhafioghjhlbpb & Cursor-A custom cursor        \\
      \rowcolor{black!10} \multicolumn{2}{l}{Volume boosters}          \\
      chmfnmjfghjpdamlofhlonnnnokkpbao & Soundboost                    \\
      hinhmojdkodmficpockledafoeodokmc & HyperVolume                   \\
      \rowcolor{black!10} \multicolumn{2}{l}{Reader mode extensions}   \\
      icnekagcncdgpdnpoecofjinkplbnocm & Easyview Reader view          \\
      dppnhoaonckcimpejpjodcdoenfjleme & Readl Reader mode             \\
      \rowcolor{black!10} \multicolumn{2}{l}{Ad blockers}              \\
      obeokabcpoilgegepbhlcleanmpgkhcp & Venus Adblock                 \\
      bkpdalonclochcahhipekbnedhklcdnp & Epsilon Ad blocker            \\
      \bottomrule
    \end{tabular}
  }
\end{table}

%% file: sections/appendix-case-studies.tex
\section{Malware Extensions From Case Studies}
\label{sec:appendix-case-studies}

Table~\ref{tab:malware-case-studies} contains relevant items from the top most
popular extensions labeled as malware by the CWS that do not appear in VirusTotal.

\begin{table}[ht!]
  \centering
  \caption{Relevant popular malware extensions that do not appear in VirusTotal.}
  \label{tab:malware-case-studies}
  \resizebox{\columnwidth}{!}{%
    \setlength{\tabcolsep}{0.5em}
    \small
    \begin{tabular}{ll}
      \toprule
      \thead[l]{Extension}             & \thead[l]{Name}                   \\
      \midrule
      \rowcolor{black!10} \multicolumn{2}{l}{Facebook stealers}            \\
      ffmdedmghpoipeldijkdlcckdpempkdi & Bookmarks Menu                    \\
      dkpedpjjafnceedhomeijlphmjbblmdj & Currency Converter PRO            \\
      jmphljmgnagblkombahigniilhnbadca & Open link in same tab, pop-up ... \\
      adkpffmlkncmmimpnmogphiijidakdhm & bilibili                          \\
      \rowcolor{black!10} \multicolumn{2}{l}{Google Tag Manager}           \\
      lfagjcmdalpklemkmdcblfghhkjjohbm & Colorize Facebook                 \\
      gahgachhcblgfnjdfghcjcpgbkbadfgg & Easy Font Changer                 \\
      lkcdbmaggddpfmfdbcloicogiaoepddk & Floating video plus               \\
      njkmonlnhpfkaldcnhikggmdaepedcep & Instant Eyedropper                \\
      jcjhgomglcabcikgghokgnheeeobakkb & La notte                          \\
      cnfianechkepmfdoakelcbamnbfbecke & Loudly                            \\
      idgifckkbacpebckkblhaopkfeikgipf & Oscura dark theme                 \\
      eemiojeoeomfggoapmnfnmpnkieojonj & PDF tools all-in-one              \\
      konkphcpahjcebjdfkeihbalppeicalp & Top Video Downloader              \\
      ailljajgcdcaadgmbncpfnofjanoabfn & Video Download Center             \\
      kdnlfofefaichijbmflgibbdlfdapmbe & Youtube Color Changer             \\
      mgccclinjajhpeiciiaflagddlhcillp & Zoom it                           \\
      \bottomrule
    \end{tabular}
  }
\end{table}